\RequirePackage{lineno}
\documentclass[aps,superscriptaddress]{revtex4-2}
\usepackage{amssymb,amsmath,epsfig}
\usepackage{subfigure}
\usepackage{color}
\usepackage[colorlinks=true, pdfstartview=FitV, linkcolor=blue, citecolor=red, urlcolor=magenta, breaklinks=true]{hyperref}

\usepackage{graphicx}
\usepackage{latexsym}
\usepackage{amsmath}
\usepackage{amsfonts}
\usepackage{amssymb}
\usepackage{booktabs}
\usepackage{multirow}
\usepackage{verbatim}
\usepackage{orcidlink}
\graphicspath{{Graf/}}
\usepackage{array}
\usepackage{float}
\usepackage{booktabs} 
\usepackage{array}

\usepackage{setspace}

\begin{document}


\title{The dark sector of the Universe as a scalar field in Horndeski Gravity}


\author{M. S. Oliveira \orcidlink{0009-0009-2742-2311}}\email{micaeloli17@gmail.com}
\affiliation{Departamento de F\'{\i}sica, Universidade Federal de Campina Grande
Caixa Postal 10071, 58429-900 Campina Grande, Para\'{\i}ba, Brazil}

\author{F. A. Brito \orcidlink{0000-0001-9465-6868}}\email{fabrito@df.ufcg.edu.br}
\affiliation{Departamento de F\'{\i}sica, Universidade Federal de Campina Grande
Caixa Postal 10071, 58429-900 Campina Grande, Para\'{\i}ba, Brazil}
\affiliation{Departamento de F\'isica, Universidade Federal da Para\'iba, 
Caixa Postal 5008, 58051-970 Jo\~ao Pessoa, Para\'iba, Brazil}

\author{J. A. V. Campos \orcidlink{0000-0002-4252-2451}}\email{a.campos@uaf.ufcg.edu.br}
\affiliation{Departamento de F\'{\i}sica, Universidade Federal de Campina Grande
Caixa Postal 10071, 58429-900 Campina Grande, Para\'{\i}ba, Brazil}


\begin{abstract}

\begin{spacing}{1}
In the present work, we study a subclass of Horndeski gravity characterized by a non-minimal derivative coupling between a scalar field and the Einstein tensor, as a possible alternative to alleviate the observational tension associated with estimates of the Hubble constant $H_{0}$. 
Two scenarios within a flat FRW spacetime were considered. In the first case, the scalar field mimics cold dark matter, whereas in the second case, it acts as dark energy. We derive the dynamical equations and perform a statistical analysis using observational data of $H(z)$, obtaining constraints for the cosmological parameters. The results indicate that the model can effectively fit the cosmic expansion rate at late epochs, providing values of $H_0$ that are more compatible with local measurements.
These results suggest that the non-minimal coupling sector in the Horndeski context constitutes a viable and promising approach to alleviate the $H_0$ tension and investigate scenarios beyond the standard cosmological model.
\end{spacing}

\end{abstract}
 \maketitle


\section{Introduction}

\label{sec:intro}
The $\Lambda$CDM concordance cosmological model, based on general relativity, associated with the cold dark matter sector and the cosmological constant, is a well-defined model that has been highly effective in describing the evolution of the Universe, accurately reproducing observations such as the cosmic microwave background and the distribution of large structures \cite{Bull2016}. 
However, the $\Lambda$CDM model presents theoretical and phenomenological limitations that challenge its consistency and motivate new investigations \cite{Perivolaropoulos2022}. From a theoretical point of view, fundamental questions such as the origin and nature of the cosmological constant and the lack of an explanation for dark matter remain open. 
The model also presents difficulties in explaining certain observational discrepancies, such as the tensions in the determination of the Hubble constant, the behavior of structures on small scales, and several other less discussed tensions \cite{Perivolaropoulos2022, Abdalla2022}. These limitations reinforce the need to study beyond the standard model, driving the development and exploration of alternative theories and extensions to the $\Lambda$CDM paradigm.

One of the main tensions related to $\Lambda$CDM concerns the current value of the Hubble parameter $H_{0}$. According to the most recent global measurements by the Planck Collaboration in 2020 (P20), $H_0 = (67.40 \pm 0.50)~\mathrm{km} \cdot \mathrm{s}^{-1} \cdot \mathrm{Mpc}^{-1}$ was estimated at the confidence level $68\%$ \cite{Planck2018}. In contrast, the latest local measurements performed by the SH$0$ES collaboration in 2022 (R22), based on Cepheid-calibrated supernovae, provided a significantly larger value, estimated at $H_0 = (73.04 \pm 1.04)~\mathrm{km} \cdot \mathrm{s}^{-1} \cdot \mathrm{Mpc}^{-1}$ at confidence level $68\%$, representing a discrepancy of about $5.0 \sigma$ between both $H_{0}$ values \cite{Riess2022}.
Although extensive discussions have sought to determine whether this tension can be attributed to unidentified systematic errors, several observations with other alternative methods have also indicated a tension in $H_{0}$ inferred from $\Lambda$CDM \cite{Freedman2021, Anand2022, deJaeger2022, Pesce2020, Shajib2023}.
Thus, there is growing evidence that the discrepancy in the $H_{0}$ values may indeed be an indication of new physics beyond the Standard Model.  
Recently, several studies have tried to solve this question, using different methods \cite{Das:2013sca,Verde2019, Knox2020, DiValentino2021, Perivolaropoulos2022, Abdalla2022, Zumalacarregui2020, Shah2021, Schoneberg2022, Das:2023rvg}. Since the tension in $H_{0}$ suggests a faster expansion of the universe than predicted by $\Lambda$CDM, a promising approach to mitigate this discrepancy involves the use of a modified gravity theory.
Such a theory should qualitatively induce a reduction in gravitational intensity during the middle and late epochs of the cosmic expansion \cite{Petronikolou2022, Petronikolou2023, Banerjee2023}.

In recent years, several studies have investigated specific subclasses of Horndeski theories in an effort to reconcile conflicting measurements of the Hubble constant $H_{0}$ \cite{Petronikolou2022, Petronikolou2023, Banerjee2023, Tiwari2024}. Horndeski gravity is regarded as one of the most general scalar-tensor theories that preserve second-order equations of motion, thereby avoiding dynamical instabilities. This framework allows for the inclusion of a scalar field that interacts directly with gravity, providing the flexibility needed to modify the Universe expansion dynamics across different cosmological epochs. In this work, we explore the application of Horndeski gravity \cite{Horndeski1974}, focusing on a specific subclass known as the John sector, which is a non-minimal derivative coupling between the scalar field and the Einstein tensor, by adopting an approach that aims to mitigate the observational tension around $H_{0}$. Two scenarios are considered. In the first, dark matter is described by a scalar field, as discussed in \cite{RINALDI201714} and  in the second, the scalar field plays the role of dark energy. Through this, estimates are possible for the values of some important cosmological quantities, such as the value of $H_0$ itself and the current matter energy density parameter $\Omega_m$.

This sector of Horndeski gravity allows a specific interaction between the scalar field and the spacetime curvature, enabling variations in the strength of gravity over time. Such behavior can contribute to the accelerated expansion of the Universe without relying exclusively on the cosmological constant. 
In recent years, this sector of theory has been widely investigated in some cosmological contexts, covering several areas of interest, such as the construction of domain wall solutions \cite{Santos2024}, simulations involving dark energy \cite{Santos2019}, and dark matter \cite{RINALDI201714, CASALINO2019100243}. 
Moreover, numerous studies have been carried out in the context of black holes, including locally asymptotically AdS and planar solutions \cite{Anabalon2014, Cisterna2014}, rotating black holes with probe strings \cite{Santos2020}, and thermodynamic analyzes of static solutions \cite{Feng2015}. In the realm of compact objects, neutron star configurations have been developed, even accounting for slow rotation \cite{Cisterna2015, Cisterna2016}. Finally, this sector has also been explored in the search for braneworld solutions \cite{Brito2020, Brito2022, Liu2016}. Thus, this approach has proven to be both promising and relevant in the literature, being a possible viable path toward alleviating the Hubble tension, by adjusting the late-time expansion rate of the universe and bringing it into better agreement with both local and global measurements of $H_{0}$.

The paper is organized as follows. In Sec. \ref{sec2} we briefly review Horndeski gravity, its field equations, and some conditions for its cosmological viability.
In Sec. \ref{sec3} we address the non-minimal coupling sector between the cosmological solutions of this class of Horndeski theories. To do so, we start from the flat FRW metric and find the differential equations to be solved in this dynamic.
Next, in Sec. \ref{sec4} we present the numerical and statistical analysis performed on the investigated model, obtaining statistical results in relation to some important cosmological quantities, in addition to the free parameters of the model, addressing issues such as a possible alternative for the relief of the Hubble tension.
In Sec. \ref{sec5} we address the issues of Laplacian instabilities and ghosts in the model. Finally, in Sec. \ref{sec6} we have the conclusions.

\section{Horndeski Gravity} \label{sec2}

In this section, we briefly discuss the Horndeski gravitational theory, introduced into the literature in 1975 by Gregory Horndeski \cite{Horndeski1974}. In recent years, this theory has been widely applied in several studies involving modified gravity, particularly in topics related to cosmology. This theory was recently rediscovered in the context of generalizations of the Galileon models and represents the most general action for a scalar-tensor theory in a four-dimensional spacetime, with second-order field equations.
\subsection{The Lagrangian}
The Horndeski theory is characterized by the action
\begin{equation} \label{acaohorndeski}
    S_{H}[g_{\mu \nu}, \phi] = \int d^{4}x \sqrt{-g} \, \mathcal{L}_{H}[g_{\mu \nu}, \phi],
\end{equation}
where $g$ is the determinant of the metric $g_{\mu \nu}$ and $\mathcal{L}_{H}$ is the Horndeski Lagrangian \cite{Horndeski1974, kobayashi2011generalized, Horndeski2024}, given by
\begin{equation} \label{eqlagHorn}
    \mathcal{L}_{H} = \sum_{i=2}^{5} \mathcal{L}_{i},
\end{equation}
where
\begin{align}
    \mathcal{L}_2 &= G_{2}(\phi, X), \label{eq:1} \\
    \mathcal{L}_3 &= -G_{3}(\phi, X) \Box \phi, \label{eq:2} \\
    \mathcal{L}_4 &= G_{4}(\phi, X) R + G_{4,X} \left[ (\Box \phi)^2 - (\nabla_{\mu} \nabla_{\nu} \phi) (\nabla^{\mu} \nabla^{\nu} \phi) \right], \label{eq:3} \\
    \mathcal{L}_5 &= G_{5}(\phi, X) G_{\mu \nu} (\nabla^{\mu} \nabla^{\nu} \phi) - \frac{1}{6} G_{5,X} \big[ (\Box \phi)^3 - 3 (\Box \phi) (\nabla_{\mu} \nabla_{\nu} \phi) (\nabla^{\mu} \nabla^{\nu} \phi) \nonumber \\
    &\quad + 2 (\nabla^{\mu} \nabla_{\alpha} \phi) (\nabla^{\alpha} \nabla_{\beta} \phi) (\nabla^{\beta} \nabla_{\mu} \phi) \big]. \label{eq:4}
\end{align}
Here, $G_{i} \, (i=2,3,4,5)$  are arbitrary functions of the scalar field $\phi$ and their canonical kinetic term $X\equiv-\frac{1}{2}\nabla^{\mu}\phi \nabla_{\mu} \phi$, with $\Box \phi=\nabla_{\mu} \nabla^{\mu} \phi$ and partial derivatives $G_{j,X}(\phi, X)=\partial G_{j}(\phi, X)/\partial X$ with $j=4,5$, $R$ is the Ricci scalar and $G_{\mu \nu}$ is the Einstein tensor. Thus, the complete action of the Horndeski theory is written as
\begin{equation} \label{acaohorndeskigeral}
    S[g_{\mu \nu}, \phi]=\int d^{4}x\sqrt{-g}(\mathcal{L}_{H}+\mathcal{L}_{M}),
\end{equation}
with $\mathcal{L}_{M}$ representing the matter and radiation content of the universe, which corresponds to a perfect fluid with power density $\rho_M$ and pressure $p_M$.

\subsection{Background Equations of Motion}

The next step is to consider an expanding Universe with a spatial flat, homogeneous, and isotropic geometry, described by the Friedmann-Robertson-Walker (FRW) metric system, written in the form
\begin{equation}\label{eqmetfrw}
    ds^2=-dt+a^2(t)\delta_{ij}dx^{i}dx^{j},
\end{equation}
from which the Friedmann equations assume the form \cite{DeFelice2012},
\begin{equation} \label{eqFRI1}
\begin{aligned}
& 2 X G_{2, X}-G_2+6 X \dot{\phi} H G_{3, X}-2 X G_{3, \phi}-6 H^2 G_4+24 H^2 X\Big(G_{4, X}+X G_{4, X X}\Big) -12 H X \dot{\phi} G_{4, \phi X}\\
&-6 H \dot{\phi} G_{4, \phi}+2 H^3 X \dot{\phi}\Big(5 G_{5, X}+2 X G_{5, X X}\Big) -6 H^2 X\Big(3 G_{5, \phi}+2 X G_{5, \phi X}\Big)=-\big(\rho_A+\rho_B\big),
\end{aligned}
\end{equation}
and
\begin{equation}\label{eqFRI2}
\begin{aligned}
G_2- & 2 X\Big(G_{3, \phi}+\ddot{\phi} G_{3, X}\Big)+2\Big(3 H^2+2 \dot{H}\Big) G_4 -12 H^2 X G_{4, X}-4 H \dot{X} G_{4, X}-8 \dot{H} X G_{4, X}\\
& -8 H X \dot{X} G_{4, X X}+2\Big(\ddot{\phi}+2 H \dot{\phi}\Big) G_{4, \phi}+4 X G_{4, \phi \phi}+4 X\Big(\ddot{\phi}-2 H \dot{\phi}\Big) G_{4, \phi X} \\
& -2 X\Big(2 H^3 \dot{\phi}+2 H \dot{H} \dot{\phi}+3 H^2 \ddot{\phi}\Big) G_{5, X}-4 H^2 X^2 \ddot{\phi} G_{5, X X}+4 H X\Big(\dot{X}-H X\Big) G_{5, \phi X}\\
& +2\left[2\Big(\dot{H} X+H \dot{X}\Big) +3 H^2 X\right] G_{5, \phi}+4 H X \dot{\phi} G_{5, \phi \phi}=-\big(p_A+p_B\big).
\end{aligned}
\end{equation}
The subscripts $A$ and $B$ in densities and pressures represent two perfect fluids, which are generally attributed to amounts of matter and radiation, and both equations \eqref{eqFRI1} and \eqref{eqFRI2} can be written in their usual forms, as follows:
\begin{equation} \label{eqfri1}
    H^2=\frac{8 \pi G}{3}\big(\rho_M+\rho_R +\rho_{\phi}\big),
\end{equation}
and
\begin{equation} \label{eqfri2}
    2\dot{H}+3H^2=-8\pi G\big(p_M+p_R+p_{\phi} \big).
\end{equation}
The modification terms of the theory can all be compactly expressed in the energy density and pressure quantities associated with the scalar field in Eqs. \eqref{eqfri1} and \eqref{eqfri2}. We vary the action \eqref{acaohorndeskigeral} with respect to $\phi$, thus obtaining the evolution of the scalar field given in the form \cite{DeFelice2012},
\begin{equation} \label{eqfri3}
    \frac{1}{a^3}\frac{d}{dt}\big(a^3 J\big)=P_{\phi},
\end{equation}
where 
\begin{equation}
\begin{aligned}
J= & \dot{\phi} G_{2, X}+6 H X G_{3, X}-2 \dot{\phi} G_{3, \phi}+6 H^2 \dot{\phi}\Big(G_{4, X}+2 X G_{4, X X}\Big)-12 H X G_{4, \phi X} \\
& +2 H^3 X\Big(3 G_{5, X}+2 X G_{5, X X}\Big)-6 H^2 \dot{\phi}\Big(G_{5, \phi}+X G_{5, \phi X}\Big) \\
\end{aligned}
\label{eqJ}
\end{equation}
and
\begin{equation}
\begin{aligned}
P_{\phi}= & G_{2, \phi}-2 X\left(G_{3, \phi \phi}+\ddot{\phi} G_{3, \phi X}\right)+6\left(2 H^2+\dot{H}\right) G_{4, \phi}\\
& +6 H\left(\dot{X}+2 H X\right) G_{4, \phi X}-6 H^2 X G_{5, \phi \phi}+2 H^3 X \dot{\phi} G_{5, \phi X},
\end{aligned}
\label{eqPf}
\end{equation}
being $J$ a current and $P_{\phi}$ the scalar source. In the equations presented above, the two perfect fluids satisfy the following continuity equations for matter and radiation, respectively,
\begin{equation}
    \dot{\rho}_M+3H\rho_M \big(1+\omega_M \big)=0,
\end{equation}
and
\begin{equation}
    \dot{\rho}_R+3H\rho_R \big(1+\omega_R \big)=0.
\end{equation}
Through this, with the set of solutions provided by these equations listed so far, it is possible to obtain the full background evolution of the universe.

\subsection{Perturbations, Instabilities and Gravitational Waves Constraints}

In the context of perturbations \cite{DeFelice2010, DeFelice2012}, the interest is focused on scalar and tensor perturbations, verifying the conditions of absence of ghost and Laplacian instabilities, which, if satisfied, will guarantee the cosmological viability of the proposed model.
In particular, for Horndeski theory to be free of Laplacian instabilities associated with propagation speed of the scalar and tensor (gravitational waves) fields, we must, respectively,  have the following expressions
\begin{equation} \label{eqcs2}
c_S^2 \equiv \frac{3\big(2 w_1^2 w_2 H-w_2^2 w_4+4 w_1 w_2 \dot{w}_1-2 w_1^2 \dot{w}_2\big)-6w_1^2\big[(1+\omega_A)\rho_A +(1+\omega_B)\rho_B\big]}{w_1\big(4 w_1 w_3+9 w_2^2\big)} \geq 0,
\end{equation}
and
\begin{equation}\label{eqcT}
c_T^2 \equiv \frac{ w_4}{w_1} \geq 0,
\end{equation}
while for the absence of ghost instabilities associated with the kinetic energy of the scalar and tensor perturbations, we must have, respectively,
\begin{eqnarray}\label{eqqs}
Q_S \equiv \frac{w_1\big(4 w_1 w_3+9 w_2^2\big)}{3 w_2^2}>0 \quad \text{and} \quad Q_T \equiv \frac{w_1}{4}>0 .
\end{eqnarray}
In the Horndeski gravity formalism applied to a FRW cosmological background, these coefficients $w_{i}\,(i=1,2,3$ and $ 4)$ are obtained from the perturbed Lagrangian, and their general expression can be found in \cite{DeFelice2012}.

The propagation of tensor perturbations, Eq. \eqref{eqcT}, is of fundamental interest, as it imposes serious constraints on Horndeski gravity, particularly on the functions $G_4(X, \phi)$ and $G_5(X, \phi)$, with respect to the speed of gravitational waves $(c_T)$. According to the GW170817 observation, conducted by the LIGO/Virgo collaboration \cite{GW170817}, together with its electromagnetic counterpart GRB 170817A \cite{Goldstein2017, Abbott2017, Abbott_2017} — in which a gravitational wave signal originating from the merger of binary neutron stars was detected — a strict bound was imposed on $c_T$ \cite{PhysRevLett.119.251301, PhysRevLett.119.251302, PhysRevLett.119.251304}, indicating that the speed of gravitational waves in the late universe must satisfy
\begin{equation}
    |c_{T}^2-1|\lesssim 10^{-15}\label{ct_cond}.
\end{equation}
This constraint implies that the speed of gravitational waves must be practically identical to that of electromagnetic waves ($c_T = c$). This becomes particularly relevant because the arbitrariness in the functions $G_i$ is drastically reduced, especially in the terms $G_4(X, \phi)$ and $G_5(X, \phi)$. In many Horndeski models, $G_5(X, \phi)$ is simply neglected in the description of the current accelerated expansion of the Universe, as well as the terms proportional to $G_{4,X}$, $G_{5,X}$, and $G_{5,\phi}$ \cite{PhysRevLett.119.251301, PhysRevLett.119.251302, PhysRevLett.119.251304, PhysRevLett.119.251303, saridakis2021modified, PhysRevD.97.104038, PhysRevD.97.061501, Matsumoto2018}. The vanishing of these terms becomes necessary to ensure that $c_T^2 = 1$ in Eq.~\eqref{eqcT}. However, there are well-established limits where it is still possible to consider $G_{4,X} \neq 0$ and $G_{5,\phi} \neq 0$ without violating the constraints imposed on $c_T$, as done in \cite{Gong2018}, for a model with non-minimal derivative coupling between the scalar field and the Einstein tensor, showing that a wide range of values for the coupling parameter $\eta$ is compatible with the stringent bounds on $c_T$. Indeed, there are also other possibilities, such as in  `extended Horndeski gravities', where this constraint can be easily respected. For a recent investigation on this issue, see, e.g., Ref.~ \cite{Santos:2024ynr}.

It is important to emphasize that propagation of signals such as GW170817 occurs on scales smaller than the homogeneity scale, although in our work we have a homogeneous cosmological background for the velocity of gravitational waves. In principle, inhomogeneities in the matter distribution could, in modified gravity theories such as Horndeski's, introduce local changes in the propagation velocity. However, as discussed in~\cite{copeland2019dark} strict constraints imposed by GW170817 and its optical counterpart GRB170817A on the velocity of gravitational waves remain valid even when considering propagation in an inhomogeneous background, since any significant deviations would be detectable.

\section{The Non-minimal Derivative Coupling Sector} \label{sec3}

In this model of Horndeski theory, our main interest lies in the non-minimal coupling between the derivative term of the scalar field and the Einstein tensor $G_{\mu \nu}$. These models are widely applied in cosmological scenarios, especially within the framework of Horndeski theories, which employ this and other more general couplings between the scalar field and curvature terms \cite{PhysRevD.81.083510, PhysRevD.79.123013, PhysRevD.104.023530, Harko2017}. They are also explored in other extended gravity theories such as, for example, those with additional coupling to the four-dimensional Gauss–Bonnet invariant, giving rise to interesting cosmological phases \cite{granda2012dark, L.N. Granda_2012}. Moreover, this coupling has been investigated in the context of the curvature model \cite{Feng2014}, in discussions of inflation during the rapid oscillation of a scalar field \cite{Sadjadi2014}, and finally, within the framework of four-dimensional $N=1$ minimal supergravity \cite{Farakos2012}.

\subsection{The Model}

In our model, we consider the sector of Horndeski gravity that can be obtained by defining the functions $G_{i}$ as follows
\begin{equation} \label{eqmodel}
\begin{aligned}
    G_{2}(\phi, X) = \alpha X -2\kappa \Lambda - V(\phi), \qquad G_{3}(\phi, X) = 0, \\ G_{4}(\phi, X) = \kappa \qquad \text{and} \qquad G_{5}(\phi, X) = -\frac{1}{2} \eta \phi.
\end{aligned}
\end{equation}
This work investigates a model belonging to a subclass of Horndeski gravity, commonly known as the John sector \cite{Starobinsky2016, Bruneton2012}, as framed within the Fab Four (F4) formulation \cite{Charmousis2012, Charmousis2012b}. The corresponding action for the John sector is given by \cite{RINALDI201714, Santos2024, Santos2019, CASALINO2019100243, Feng2015, Anabalon2014, Santos2020, Cisterna2014, Cisterna2015, Cisterna2016, Brito2020, Brito2022},
\begin{equation} \label{eqacao}
    S[g_{\mu \nu}, \phi] = \int d^4x \sqrt{-g} \left[ \kappa( R-2\Lambda) - \frac{1}{2} (\alpha g_{\mu \nu} - \eta G_{\mu \nu}) \nabla^{\mu} \phi \nabla^{\nu} \phi - V(\phi) \right] + S_m[g_{\mu \nu}],
\end{equation}
with $\kappa = (16 \pi G)^{-1}$. The parameters that control the intensity of the couplings are $\alpha$ and $\eta$, the first is the dimensionless parameter, and the second has dimension of $(mass)^{-2}$. Note that, by defining $\alpha = 1$ and $\eta = 0$ in the action above, we recovered the usual Einstein theory, with minimally coupled gravity to the scalar field $\phi$ with potential $V(\phi)$.
A key characteristic of this sector is that the propagation speed of tensor modes, $c_{T}$, is not necessarily equal to unity throughout the redshift range, which requires that the parameters of the theory be adjusted to maintain the restriction $|c_{T}^2-1|\lesssim 10^{-15}$, as detailed in a subsequent section. In contrast to other subclasses such as Kinetic Gravity Braiding \cite{deffayet2010imperfect}, where the Lagrangian structure inherently guarantees $c_{T}^{2} =1.$

\subsection{Cosmological Dynamics of the Model} 

A possible approach to this topic consists in varying the action in Eq. \eqref{eqacao} with respect to the metric and the scalar field, thus obtaining the field equations of the Horndeski theory, which can be solved using the FRW metric presented in Eq. \eqref{eqmetfrw}. However, this algebraic procedure can be bypassed by directly applying the model equations \eqref{eqmodel} to the expressions \eqref{eqFRI1} and \eqref{eqFRI2}. In doing so, we obtain the two Friedmann equations in their standard forms
\begin{equation}
H^2 = \frac{8 \pi G}{3} \big(\rho_m + \rho_r + \rho_{\phi} + \rho_{\Lambda}\big),
\end{equation}
and
\begin{equation}
2\dot{H} + 3H^2 = -8\pi G \big(p_m + p_r + p_{\phi} + p_{\Lambda}\big).
\end{equation}
The equations corresponding to the energy density and pressure of the scalar field, associated with the model defined in \eqref{eqmodel}, now take the following forms
\begin{equation} \label{eqdensidphi}
\rho_{\phi}=\dfrac{\dot{\phi}^{2}}{2}\left(\alpha+9 \eta H^2 \right)+V(\phi)
\end{equation}
and
\begin{equation}\label{eqpphi}
    p_{\phi}=\frac{\alpha \dot{\phi}^2}{2}-\frac{\eta\dot{\phi}^2}{2} \big(2\dot{H}+3H^2\big) -2\eta H \dot{\phi} \ddot{\phi}+V(\phi).
\end{equation}
We take the derivative of $\rho_{\phi}$ with respect to time, thus obtaining its continuity equation for the scalar field, written in the form
\begin{equation} \label{eqdceponto}
    \dot{\rho_{\phi}}-\phi \ddot{\phi} \big(\alpha + 9\eta H^2\big) - 9 \eta  \dot{\phi}^{2} H \dot{H}-\dot{\phi} V_{\phi}(\phi)=0,
\end{equation}
while the continuity equations for the other components of matter and radiation, which in turn satisfy their usual forms, are written in the following way
\begin{equation} \label{eqdm}
    \dot{\rho}_m+3H \rho_m=0 \qquad \text{and} \qquad \dot{\rho}_r+4H \rho_r=0.
\end{equation}
Another important equation that drives the cosmological evolution is the equation of motion for $\phi$, which can be obtained directly by applying \eqref{eqmodel} to \eqref{eqfri3}, obtaining the following expression
\begin{equation} \label{eqceponto}
    \Ddot{\phi}+3H \dot{\phi}+\frac{6 \eta \dot{\phi}H \dot{H}}{\alpha +3\eta H^2}+\frac{V_{\phi}(\phi)}{\alpha +3\eta H^2}=0.
\end{equation}
For $V_{\phi}=0$, we return to the form found in \cite{RINALDI201714}. Finally, we will compute equations for the energy density parameters whose components satisfy
\begin{equation}
\Omega_m+\Omega_r+\Omega_{\Lambda}+\Omega_{\phi}=1.
\end{equation}
The cosmic time $t$, is related to the redshift $z$ by the relation $1+z=a(t_0)/a(t)$, with $t_0$ being the current time such that $a(t_0)=1$. The differential equations can be written in terms of $z$ by using the following relation
\begin{equation}\label{eqtransf}
   \frac{d}{dt}=-(1+z)H(z)\frac{d}{dz}. 
\end{equation}
With this, we rewrite the differential equations in terms of $z$. The Hubble function is expressed in the form
\begin{equation}\label{funhubble}
    H(z)= \sqrt{H_{0}^{2}\big[\Omega_{m0}(1+z)^{3} +\Omega_{r0}(1+z)^{4} +\Omega_{\Lambda 0}\big]+\dfrac{8\pi G\rho_{\phi}}{3}},
\end{equation}
where $H_0$ is the Hubble parameter today, while $\Omega_{i0}$ stands for the present values of the energy density parameters of each component, and $\Omega_{\phi}$ represents the quantity associated with the relative energy density of the scalar field. In the following sections, we will consider two specific cases for the field $\phi$.

We rewrite the equation \eqref{eqdensidphi} in terms of $z$ as follows
\begin{equation} \label{eq1.1fried}
    \rho_{\phi}=\dfrac{(1+z)^{2}\phi'^{2}H^{2}\left(\alpha+9 \eta H^2 \right)}{2}+V(\phi).
\end{equation}
In the same way, we have that equations \eqref{eqdceponto} and \eqref{eqceponto} are given respectively by
\begin{equation} \label{eqdif5}
\rho'_{\phi} + \phi (1 + z)(H'\phi' + H\phi'')(\alpha - 9\eta H^2) - 9\eta (1 + z)^2 H^2 H' \phi'^2 - \phi' V(\phi) = 0,
\end{equation}
and
\begin{equation} \label{eqdif6}
(1 + z)^2 H(H'\phi' + H\phi'') - 3(1 + z) H^2 \phi' + \frac{6\eta (1 + z)^2 H^3 H'\phi'}{\alpha + 3\eta H^2} + \frac{V_{\phi}(\phi)}{\alpha + 3\eta H^2} = 0.
\end{equation}
Combining equations (\ref{eqdif5}) and (\ref{eqdif6}), we obtain a first-order differential equation in $\phi$, given by
\begin{equation} \label{eqGG}
\begin{aligned}
(\alpha + 3\eta H^2)\rho'_{\phi} - 3(1 + z) H^2 \phi'^2 \big[ (\alpha + 3\eta H^2)(\alpha + 9\eta H^2) + \eta (1 + z) H H'(\alpha - 9\eta H^2) \big] \\ + 6\eta H^2 \phi' V'(\phi) = 0.
\end{aligned}
\end{equation}

\section{Numerical and statistical analysis} \label{sec4}
In this section, we describe the procedures performed for the numerical and statistical analysis of the cosmological background given by the equations \eqref{eqmodel}. The equations of motion of the model are solved numerically using the residual method implemented in the Mathematica software. In this way, we obtain the solutions of the differential equations involved. For statistical analysis, we employ the Markov Chain Monte Carlo (MCMC) method, which determines the parameter space for the free parameters in the model.
In this way, we obtain some relevant quantities, such as the Hubble parameter $H(z)$ and the energy density parameters $\Omega_i$. 

We apply Bayesian sampling of the posterior probability distribution of these parameters, performing an MCMC analysis implemented in the Wolfram Mathematica software, obtaining the best fitting constraints for the following free parameters: $\alpha$, $\eta$, $H_0$ and $\Omega_{i 0}$. This is done by confronting the Hubble function with experimental data obtained through the Cosmic Chronometer (CC), Baryon Acoustic Oscillations (BAO), and the SH$0$ES Collaboration methods for measurements of $H(z)$ at low redshifts, as we specify below.
The agreement between the results of $H(z)$ from numerical integration and the observational data is evaluated with the following chi-square function
\begin{eqnarray}
\chi^{2} = \sum_{i=1}^{N}\dfrac{\big[H(x,z_{i}) - H_{obs}(z_{i})\big]^{2}}{\sigma^{2}(z_{i})},
\end{eqnarray}
where $N$ is the maximum number of observational data and $x$ in the $H$ function are the free parameters of the model. In our analyses, we employ an exponential-type scalar potential, expressed in the form
\begin{equation}
    V(\phi)=V_0 e^{-\lambda\phi}.
\end{equation}
The exponential potential has been long well-justified in the literature. For recent motivations, see, for example, in dilatonic \cite{Brito:2023bkc, Brito:2025fmj}, quintessence  \cite{Akrami:2025zlb} and deformed Starobinsky  \cite{SantosdaCosta:2020dyl} models.

We establish initial conditions for the scalar field and potential amplitude $V_{0}$, and thus numerically solve the set of differential equations given by the equations \eqref{eq1.1fried} and \eqref{eqGG}. In this paper, we consider two cases for this model: In the first case, the scalar field plays the role of dark matter, while in the second case, we attribute to the scalar field the description of the amount of dark energy.

\subsection{About $H(z)$ data}

Here we discuss the observational datasets of $H(z)$ used to fit the model.
We used the expansion rate value at redshift $z=0$, one of the latest results obtained for the Hubble constant, estimated at $H_0 = 73.04 \pm 1.04~(\mathrm{km/s/Mpc})$. This value comes from the Hubble Space Telescope (HST) measurement together with the Supernova Collaboration $H_0$ for the Equation of State (SH$0$ES) \cite{Riess2022}.
Regarding the baryon acoustic oscillation (BAO) measurements of $H(z)$ from the Sloan Digital Sky Survey Collaboration (SDSS), we use the dataset specified and cited in \cite{Planck2018}, where we can find them in references \cite{Alam2017, Zarrouk2018, Agathe2019, Blomqvist2019}.
For cosmic chronometer (CC) data, we used a list of measurements with 33 results of $H(z)$, which can be found in Table III of \cite{PhysRevD109_023525}, together with their respective references for each point.
In the results shown below, we made some combinations between the aforementioned datasets.

\subsection{Case I: scalar field as dark matter}
For this case, we have the contribution of the cosmological constant $\Lambda$ playing the role of dark energy, while the scalar field plays the role of dark matter, and thus the quantity $\Omega_{m}$ represents only the baryonic matter. In addition, we have the non-minimal coupling parameters of the model $\alpha$ and $\eta$. 
For statistical analysis using the following values for the potential parameters $V_{0}=15.0 \times 10^{-124}\,M_{\mathrm{Pl}}^{4}$ and $\lambda = 0.28$, we assume a small initial value for $\phi'(z=0) = 0.5\times10^{-6}\, M_{\mathrm{Pl}}$ such that it guaranties condition \eqref{ct_cond}, taking into account equations \eqref{eqcT}, \eqref{eqw1}, and \eqref{eqw4}. Furthermore, the initial condition $\phi(z=0)$ can be obtained so that it satisfies \eqref{eq1.1fried} for $z=0$ using the aforementioned values for the potential parameters and $\phi'(z=0)$.
We present in Table \ref{tabss} the best fit (with $1\sigma$ constraints) for the free parameters of the model, while in Table \ref{tabvc} we have the mean values (with $2\sigma$ constraints) considered for this case of the non-minimal derivative coupling model of Horndeski gravity. These results were obtained by MCMC sampling, where we used different $H(z)$ data at low and intermediate redshifts.
\begin{table}[htbp]
    \centering
    \begin{tabular}{| l | l | >{\centering\arraybackslash}p{2.5cm} | >{\centering\arraybackslash}p{3cm}|}
    \hline
        \multirow{1}{*}{Model} & \multirow{1}{*}{Parameter} & 
        \multirow{1}{*}{CC} & 
        \multirow{1}{*}{CC$+$BAO$+$SH$0$ES} \\
    \hline
    \hline
    \multirow{6}{*}{$\phi$ as DM}    
         &  $\alpha (\times 10^{-2})$ & $0.250 \pm 0.012$ & $0.231 \pm 0.018$ 
         \\
         &   $\eta(\times 10^{118}) [M_{\mathrm{Pl}}^{-2}]$   & $0.981 \pm 0.019$ &  $0.992 \pm 0.020$ 
         \\
         &   $H_0[\mathrm{km} / \mathrm{s} / \mathrm{Mpc}]$  & $68.8 \pm 0.9$ & $73.5 \pm 0.6$ 
         \\
         &   $\Omega_m$      & $0.048 \pm 0.003$ &  $0.054 \pm 0.003$ 
         \\
         &   $\Omega_{\Lambda}$ & $0.674 \pm 0.008$ & $0.696 \pm 0.007$ 
         \\
    \hline
    \end{tabular}
    \caption{Table of best-fitting estimated values for the free parameters of the model where $\phi$ plays the role of dark matter, with the 68\% confidence interval.}
    \label{tabss}
\end{table}

\begin{table}[htbp]
    \centering
    \begin{tabular}{|l|l|
    >{\centering\arraybackslash}p{2cm}| >{\centering\arraybackslash}p{2cm}|
    >{\centering\arraybackslash}p{2cm}| >{\centering\arraybackslash}p{2cm}|}
    \hline
        \multirow{2}{*}{Model} & \multirow{2}{*}{Parameter} & 
        \multicolumn{2}{c|}{CC} & 
        \multicolumn{2}{c|}{CC$+$BAO$+$SH$0$ES} \\
        \cline{3-6}
        & &
       $1\sigma$ & $2\sigma$ & 
       $1\sigma$ & $2\sigma$ \\
    \hline
    \hline
    \multirow{6}{*}{$\phi$ as DM}
         & $\alpha (\times 10^{-2})$ & $0.250_{-0.012}^{+0.012}$ & $0.250_{-0.022}^{+0.025}$ & $0.231_{-0.018}^{+0.018}$ & $0.231_{-0.029}^{+0.04}$   
         \\
         & $\eta(\times 10^{118}) [M_{\mathrm{Pl}}^{-2}]$  & $0.982_{-0.019}^{+0.020}$ & $0.98_{-0.04}^{+0.04}$ & 
         $0.992_{-0.020}^{+0.020}$ & $0.99_{-0.04}^{+0.04}$    
         \\
         & $H_0[\mathrm{km} / \mathrm{s} / \mathrm{Mpc}]$ & $68.8_{-0.9}^{+0.9}$ 
         & $68.8_{-1.7}^{+1.7}$
         & $73.5_{-0.6}^{+0.6}$
         & $73.5_{-1.1}^{+1.1}$
         \\
         & $\Omega_m$ & $0.048_{-0.003}^{+0.003}$ &
         $0.048_{-0.006}^{+0.005}$ & $0.054_{-0.003}^{+0.003}$ & 
         $0.054_{-0.005}^{+0.005}$
         \\
         & $\Omega_{\Lambda}$ &
         $0.674_{-0.009}^{+0.008}$ &
         $0.674_{-0.017}^{+0.016}$ &
         $0.696_{-0.007}^{+0.007}$ &
         $0.696_{-0.015}^{+0.014}$
         \\
    \hline
    \end{tabular}
    \caption{Estimated mean values for the free parameter of the model where $\phi$ plays the role of dark matter at the 68\% and 95\% confidence intervals, where $\Omega_{m}$ describes only the baryonic part.}
    \label{tabvc}
\end{table}

We performed an individual analysis for the parameter $H(z)$, obtaining the best fit $H_{0} = 68.8\pm 0.9~(\mathrm{km/s/Mpc})$ within the confidence level $68\%$ of the CC measurements, which is in tension with R22 at $3.08\sigma$. This result is midway between the R22 value and the Planck value.
For the analysis of the combined measurements of CC$+$BAO$+$SH$0$ES, we obtain the best fit $H_{0}=73.5\pm 0.6~(\mathrm{km/s/Mpc})$ within the $68\%$ confidence level, presenting a value very close to R22 with a tension of approximately $0.38\sigma$, which is particularly interesting due to the precision level and proximity to the measured value locally.  However, it raises the tension with P20.

 \begin{figure}[htbp]
 \centering
\subfigure{\includegraphics[width=.45\textwidth]{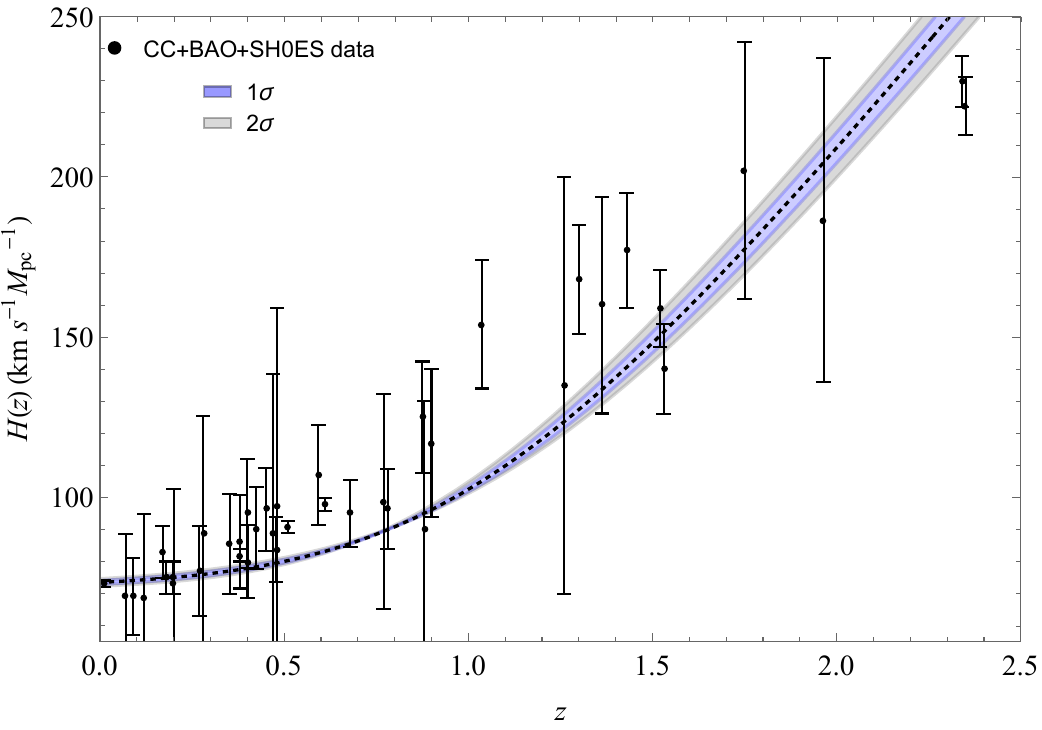}\label{grafH01lambda}}
 \quad
 \subfigure{\includegraphics[width=.45\textwidth]{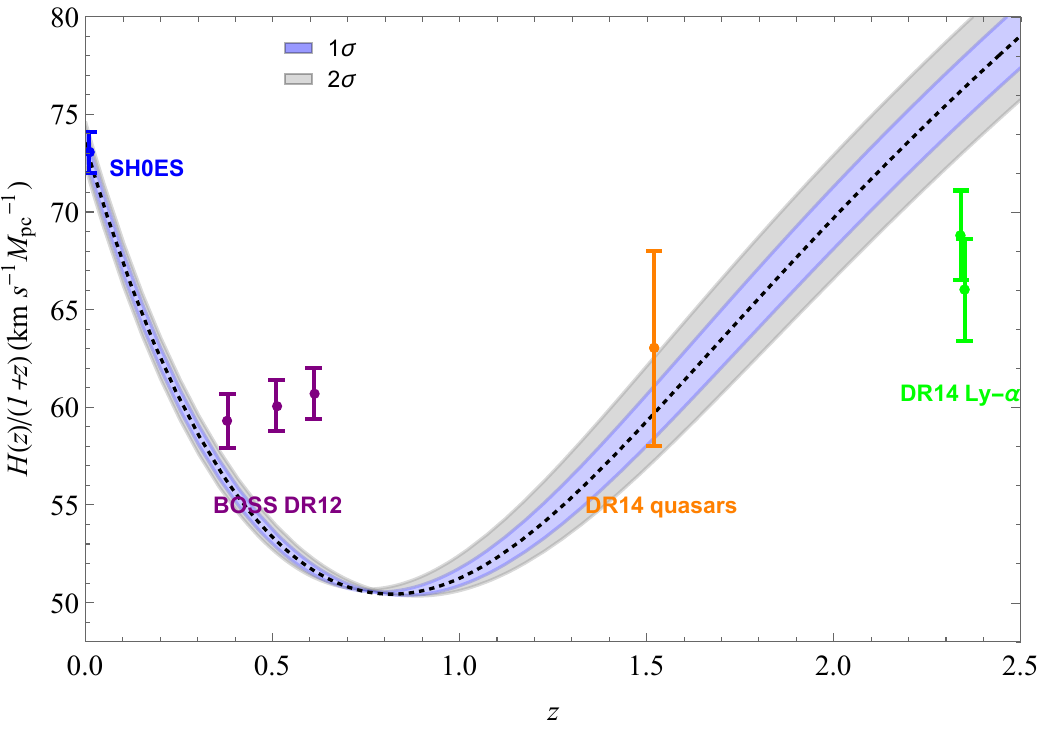}\label{grafH02lambda}}
  \caption{\footnotesize{In the left panel, we present the evolution of the Hubble parameter as a function of redshift, confronted with observational data of $H(z)$. In the right panel, we have the evolution of the normalization $H(z)/(1+z)$. In both graphs, we show the confidence bands of $1\sigma$ and $2\sigma$ results for $\phi$ as dark matter.}}
 \label{q5}
\end{figure}

\begin{figure}[htbp]
 \centering
 \includegraphics[width=.65\textwidth]{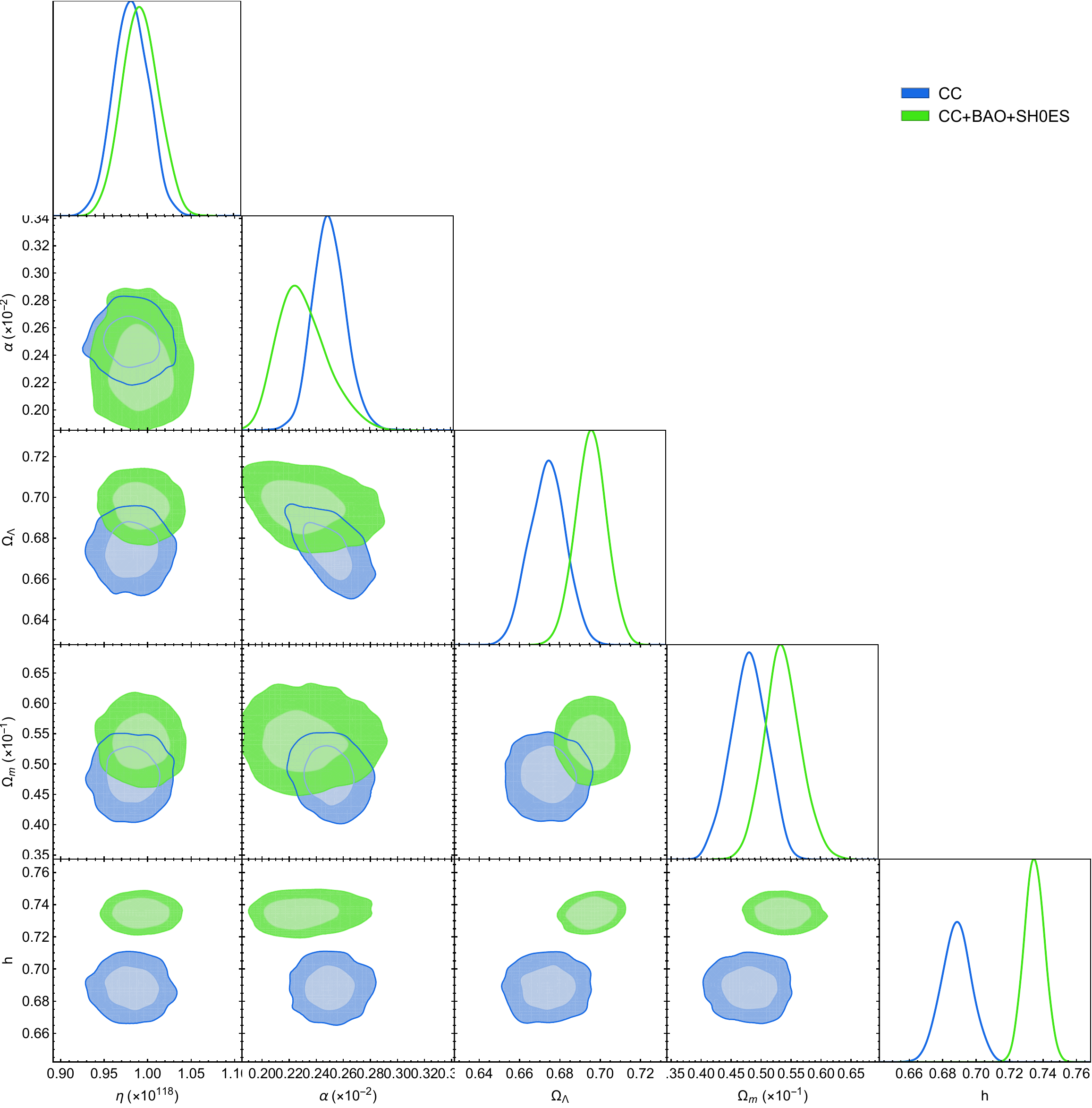}  \caption{\footnotesize{Posterior distribution of the model parameters $\alpha$ and $\beta$ and the background quantities $h$, $\Omega_m$ and $\Omega_{\Lambda}$. Here, for a better display of the graph, we use $h=H_{0}/100$. For each parameter, we have the respective contour plots with confidence regions $1\sigma$ and $2 \sigma$ of the MCMC sampling, for individual CC data (blue regions) and for combined CC$+$BAO$+$SH$0$ES data (green regions).}}
\label{grafdistcase1}
\end{figure}

Thus, we present in Fig. \ref{q5} the graphs of $H(z)$ and its normalization $H(z)/(1+z)$ with their respective confidence bands. In Fig. \ref{grafdistcase1}, we have the results of the analysis as posterior probability distribution of the model parameters $\alpha$ and $\eta$, in addition to the other parameters of the cosmological background, with their respective contour plots referring to the confidence regions $1\sigma$ and $2\sigma$ of the MCMC sampling, obtaining results for the CC data set and for the combination CC$+$BAO$+$SH$0$ES.

\subsection{Case II: Scalar field as dark energy }
In this case, we consider the scalar field $\phi$ playing the role of dark energy. Now, we have that $\Omega_{m}$ represents the contribution of dark matter and baryonic matter, while $\Omega_{\phi}$ corresponds to the amount of dark energy. The procedure is similar to that carried out for Case I, with one less free parameter. For Case II, we adopt the values of the potential parameters $V_0=10 \times 10^{-124}\, M_{\mathrm{Pl}}^{4}$ and $\lambda = 0.5$, using the initial condition $\phi'(z=0) =0.5\times 10^{-6}\, M_{\mathrm{Pl}}$. For this analysis, we have added statistics using the individual BAO data, shown in Tables \ref{tabbf2} and \ref{tabvb}

\begin{table}[htbp]
    \centering
    \begin{tabular}{| l | l | >{\centering\arraybackslash}p{2.5cm} | >{\centering\arraybackslash}p{2.5cm}| >{\centering\arraybackslash}p{3cm} |}
    \hline
        \multirow{1}{*}{Model \quad} & \multirow{1}{*}{Parameter} \quad & 
        \multirow{1}{*}{BAO} &
        \multirow{1}{*}{CC} & 
        \multirow{1}{*}{CC$+$BAO$+$SH$0$ES} \\ 
    \hline
    \hline
    \multirow{5}{*}{$\phi$ as DE}
        
         &  $\alpha$ $ (\times 10^{1})$ & $0.602 \pm 0.049$ & $0.600 \pm 0.050$ & $0.600 \pm 0.050$
         \\
          &   $\eta$ $ (\times 10^{118})$ $[M_{\mathrm{Pl}}^{-2}]$    & $0.349 \pm 0.049$ &  $0.349 \pm 0.050$ & $0.349 \pm 0.050$ 
          \\
          &   $H_0$ $[\mathrm{km} / \mathrm{s} / \mathrm{Mpc}]$  & $68.8 \pm 1.1$ & $69.1 \pm 1.4$ & $70.8 \pm 0.8$ 
    \\
   &   $\Omega_m$      & $0.285 \pm 0.016$ &  $0.296 \pm 0.018$ & $0.267 \pm 0.012$
   \\
    \hline
    \end{tabular}
    \caption{Best-fit estimates of the free parameters of the model where $\phi$ plays the role of dark energy, with a 68\% confidence interval.}
    \label{tabbf2}
\end{table}

\begin{table}[htbp]
    \centering
    \begin{footnotesize}
    \begin{tabular}{|l|l|
    >{\centering\arraybackslash}p{1.6cm}| >{\centering\arraybackslash}p{1.6cm}|
    >{\centering\arraybackslash}p{1.6cm}| >{\centering\arraybackslash}p{1.6cm}|
    >{\centering\arraybackslash}p{1.6cm}| >{\centering\arraybackslash}p{1.6cm}|}
    \hline
        \multirow{2}{*}{Model} & \multirow{2}{*}{Parameter} \quad & 
        \multicolumn{2}{c|}{BAO} & \multicolumn{2}{c|}{CC} & 
        \multicolumn{2}{c|}{CC$+$BAO$+$SH$0$ES} \\
        \cline{3-8}
        & &
       $1\sigma$ & $2\sigma $ & 
       $1\sigma$ & $2\sigma$ &
       $1\sigma$ & $2\sigma$ \\
    \hline
    \hline
    \multirow{5}{*}{$\phi$ as DE}
         & $\alpha$ $ (\times 10^{1})$ & $0.602^{+0.051}_{-0.050}$ &
         $0.602^{+0.096}_{-0.097}$ & 
         $0.600^{+0.050}_{-0.051}$ &
         $0.600^{+0.098}_{-0.098}$ &
         $0.600^{+0.050}_{-0.050}$ &
         $0.600^{+0.100}_{-0.097}$ \\
         & $\eta$ $ (\times 10^{118})$ $[M_{\mathrm{Pl}}^{-2}]$ &  
         $0.350^{+0.048}_{-0.050}$ & $0.350^{+0.099}_{-0.097}$ & $0.349^{+0.050}_{-0.050}$ & $0.349^{+0.10}_{-0.10}$ & $0.349^{+0.050}_{-0.050}$ & $0.349^{+0.096}_{-0.098}$  \\
         & $H_0$ $[\mathrm{km} / \mathrm{s} / \mathrm{Mpc}]$ & $68.8 ^{+1.1}_{-1.1}$ & $68.8 ^{+2.2}_{-2.2}$ & $69.0 ^{+1.4}_{-1.4}$ & $69.0 ^{+2.8}_{-2.7}$ & $70.8 ^{+0.7}_{-0.8}$ & $70.8 ^{+1.5}_{-1.5}$ \\
         & $\Omega_m$ & 
         $0.285^{+0.016}_{-0.015}$ &
         $0.285^{+0.031}_{-0.030}$ & $0.296^{+0.018}_{-0.018}$ & 
         $0.296^{+0.035}_{-0.040}$ & 
         $0.267^{+0.013}_{-0.012}$ & 
         $0.267^{+0.025}_{-0.023}$ \\
    \hline
    \end{tabular}
    \caption{Mean values estimated for the free parameters of the model where $\phi$ plays the role of dark energy, with confidence intervals of 68\% and 95\%.}
    \label{tabvb}
    \end{footnotesize}
\end{table}

We perform the same individual analysis for the parameter $H(z)$, where we obtain the best fit $H_{0} = 70.8 \pm 0.8~(\mathrm{km/s/Mpc})$ within the confidence level of $68\%$, with points of $H(z)$ from the combined CC$+$BAO$+$SH$0$ES measurements, which is in tension with R22 at $1.7\sigma$, is particularly interesting since it is much smaller than the tension existing between R22 and P20 of $5\sigma$, in addition to several other dark energy models, which provides considerable relief in the Hubble tension with the model based on Horndeski gravity.  Regarding the average values presented in Table \ref{tabvb}, all remain close to this restriction.

\begin{figure}[htbp]
 \centering
\subfigure{\includegraphics[width=.45\textwidth]{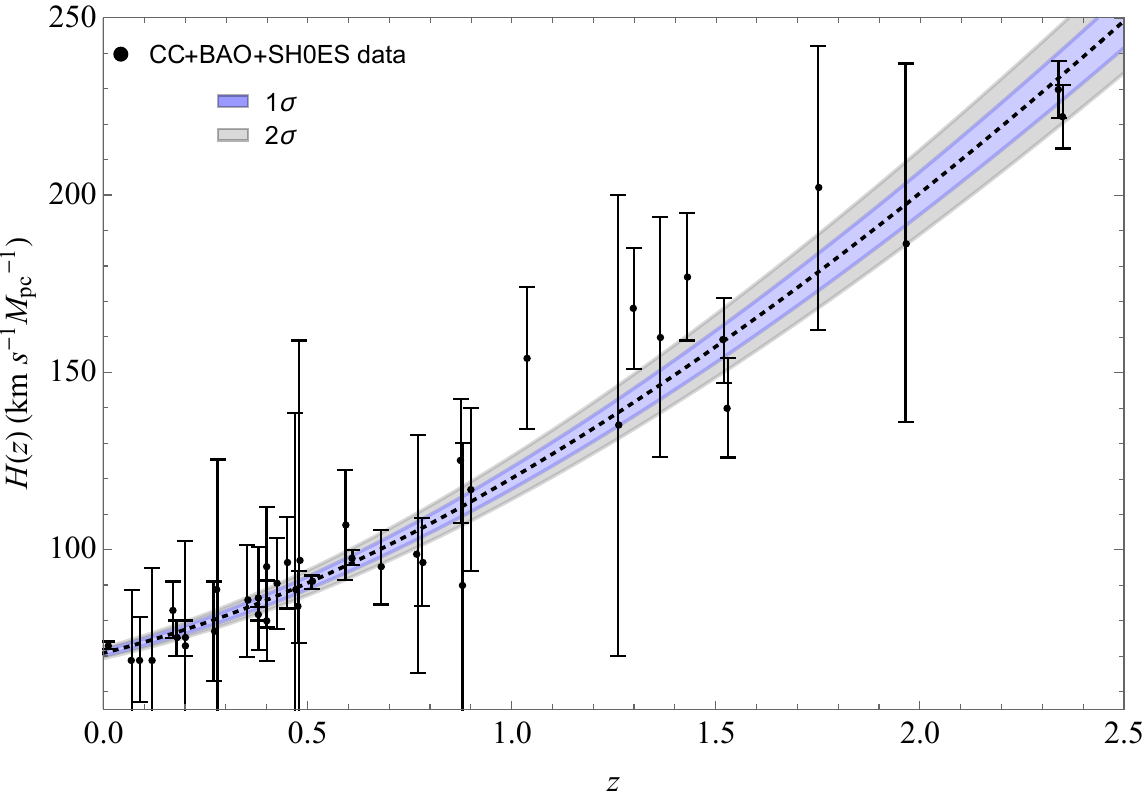}\label{grafPlotH0_CBS}}
 \quad
 \subfigure{\includegraphics[width=.45\textwidth]{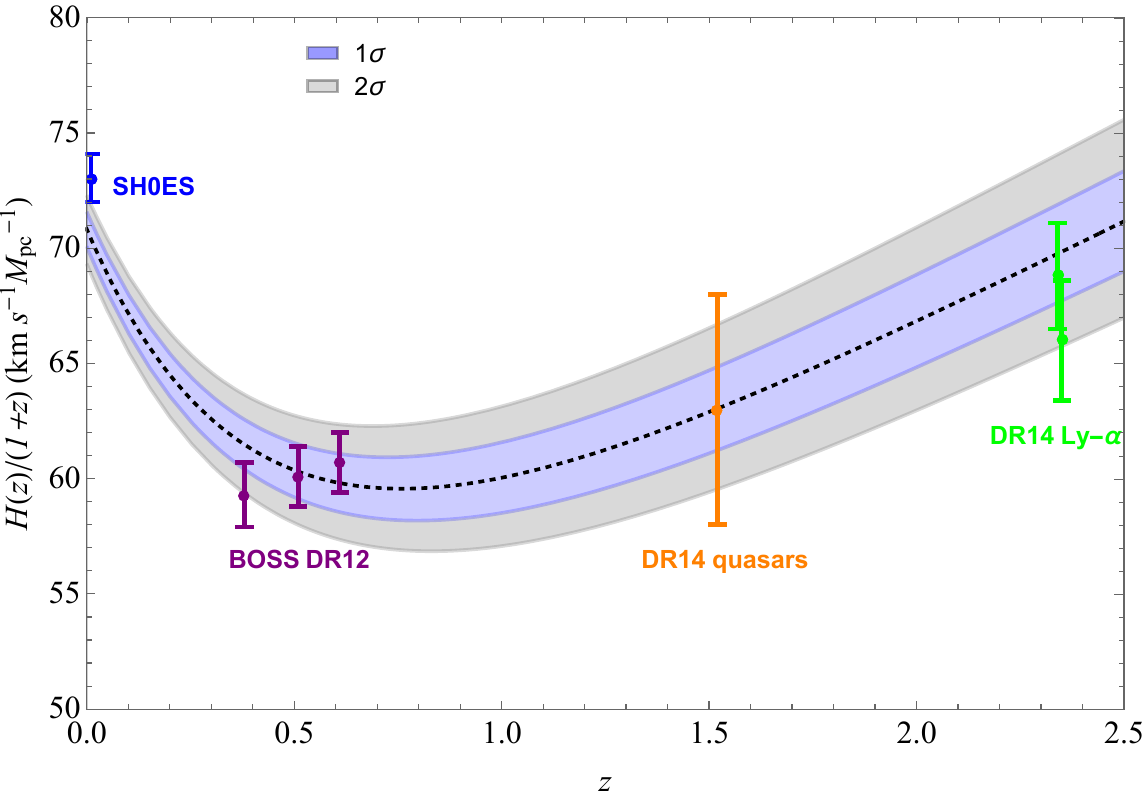}\label{grafPlotH0_CBS2}}
  \caption{\footnotesize{ In the left panel we have the evolution of the Hubble parameter as a function of redshift, compared with observational data of $H(z)$. In the panel on the right, we have the evolution of the normalization $H(z)/(1+z)$. In both graphs, we show the confidence bands of $1\sigma$ and $2\sigma$ with $\phi$ as dark energy.}}
 \label{PLOH0}
\end{figure}
\begin{figure}[htbp]
 \centering
 \includegraphics[width=.6\textwidth]{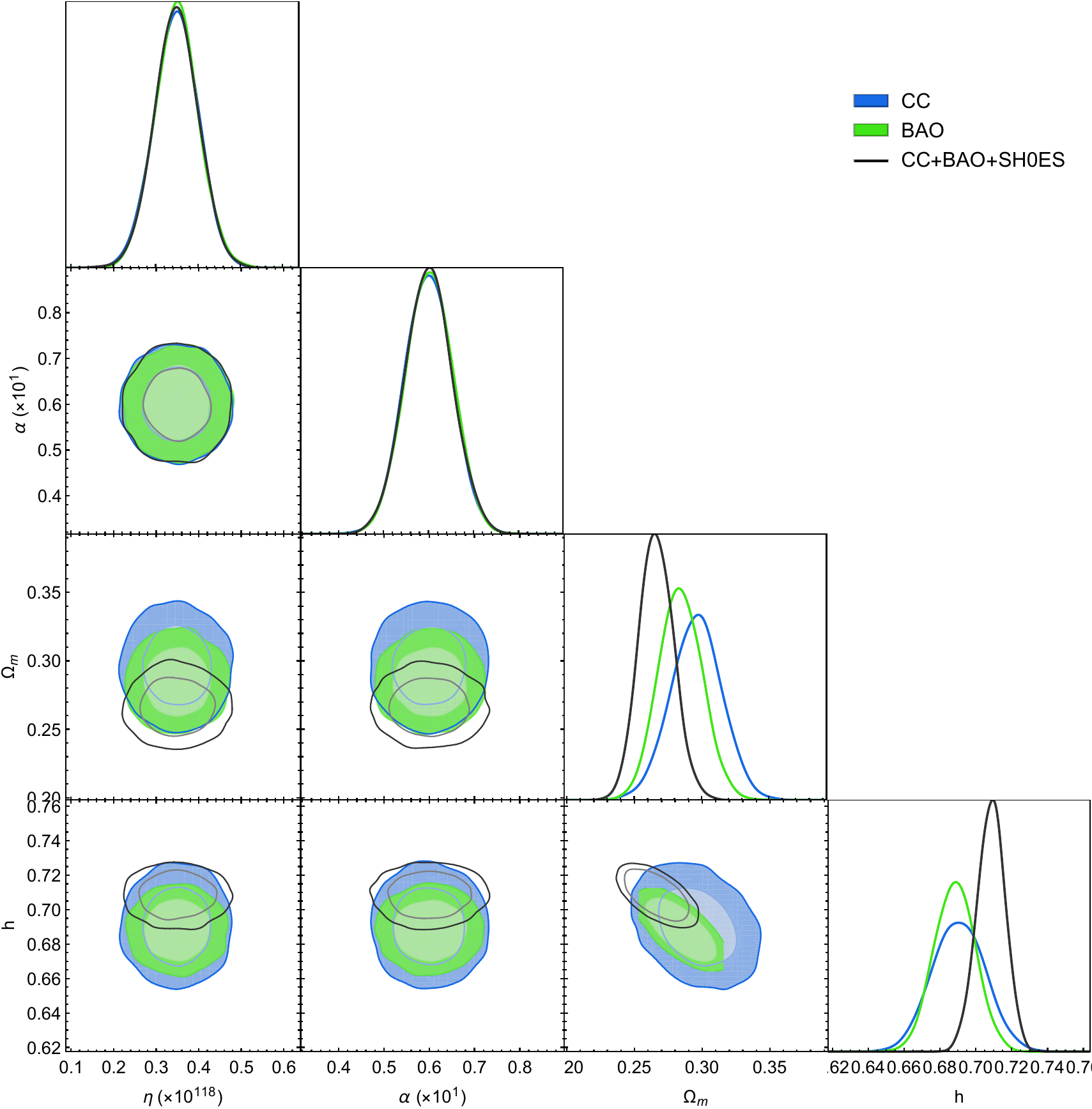}\label{grafPlotCC_BAO_SH0ES}
  \caption{\footnotesize{Posterior distribution of the model parameters $\alpha$ and $\eta$ together with the background quantities $h$ and $\Omega_m$, again we have that $h=H_{0}/100$. For each parameter, we have its respective contour plot with 1$\sigma$ and 2$\sigma$ confidence regions from the MCMC sampling, for individual CC (blue regions) and BAO (green regions) data, and also for combined CC$+$BAO$+$SH$0$ES data (black edges).}}
  \label{gju}
\end{figure}

Thus, in Fig. \ref{PLOH0} we present the graphs of $H(z)$ and its normalization $H(z)/(1+z)$ with their respective confidence bands. In Fig. \ref{gju}, the results of the analysis as posterior probability distribution of the parameters $\alpha$ and $\eta$ of the model, in addition to the other cosmological quantities, together with their respective contour plots with confidence regions 1$\sigma$ and 2$\sigma$ of the MCMC sampling.
For this case, we analyzed the individual data set of BAO and CC, and finally the combination of the data CC$+$BAO$+$SH$0$ES.

Other approaches based on Horndeski gravity aim to treat, for example, the Hubble tension, in addition to cosmological and astrophysical approaches \cite{Petronikolou2022, Banerjee2023, Tiwari2024}. The model discussed here presents flexibility in adjusting the fundamental cosmological quantities and the parameters $\alpha$ and $\eta$ of the model, producing estimates with good agreement with the standard model, allowing to better accommodate the discrepancies between local and global measurements of $H_0$.
Thus, the numerical and statistical results obtained indicate that the non-minimal derivative coupling model of Horndeski gravity has promising potential in the description of cosmological dynamics, in addition to treating the issue of the tension in $H_0$, reconciling the different observational regimes of cosmic expansion.
\section{Stability Conditions for the Model} \label{sec5}
We apply the relations of the functions $G_{i}$ expressed in \eqref{eqmodel} to obtain the equations of $w_{i}$ for the model, written in the form 

\begin{align}
w_1 &= \frac{4\kappa-\eta (1+z)^2 H^2 \phi'^2}{2}, \label{eqw1} \\
    w_2 &= 4\kappa H-3\eta (1+z)^2 H^3 \phi'^2, \label{eqw2} \\
    w_3 &= \frac{3}{2}\alpha (1+z)^2H^2\phi'^2-18 \kappa H^2 +27\eta (1+z)^2H^4\phi'^2, \label{eqw3} \\
    w_4 &= \frac{4\kappa +\eta (1+z)^2 H^2 \phi'^2}{2}. \label{eqw4}
\end{align}

\begin{figure}[htbp]
 \centering
\subfigure[]{\includegraphics[width=.45\textwidth]{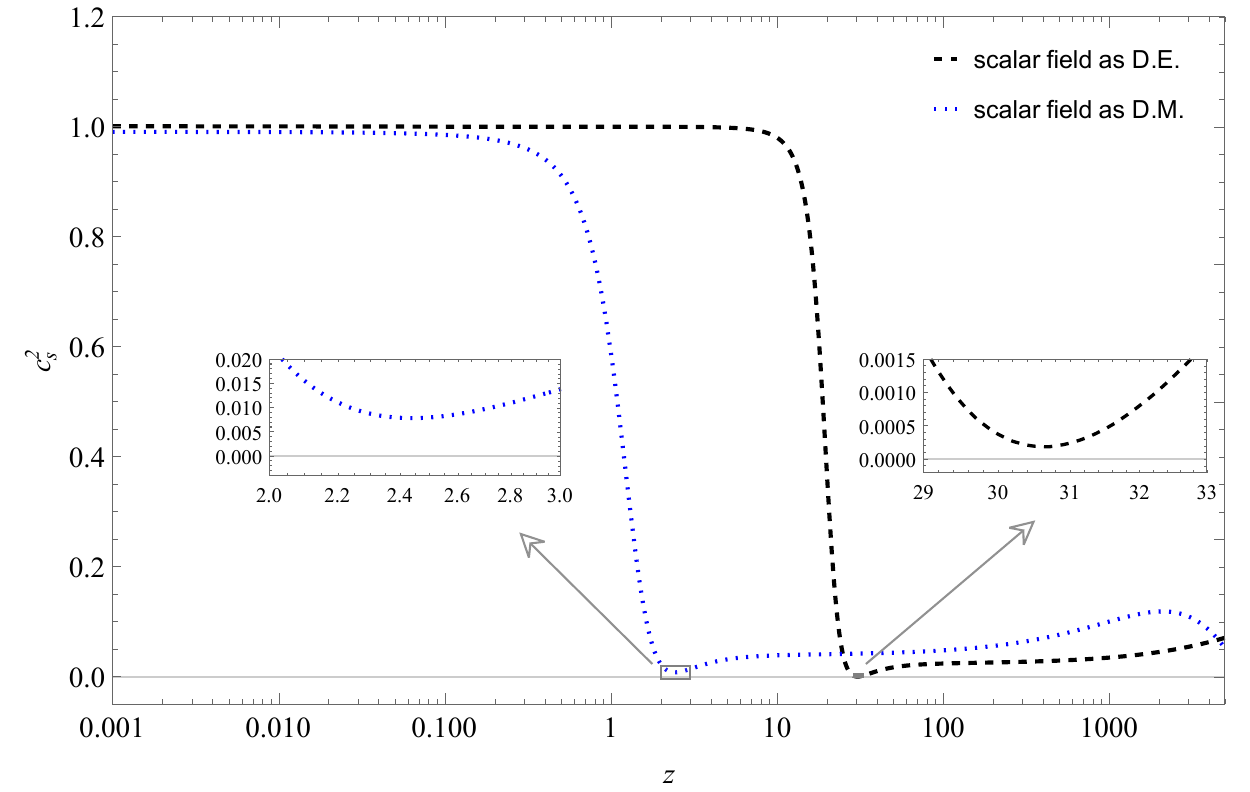}\label{grafcS}}
 \quad
 \subfigure[]{\includegraphics[width=.45\textwidth]{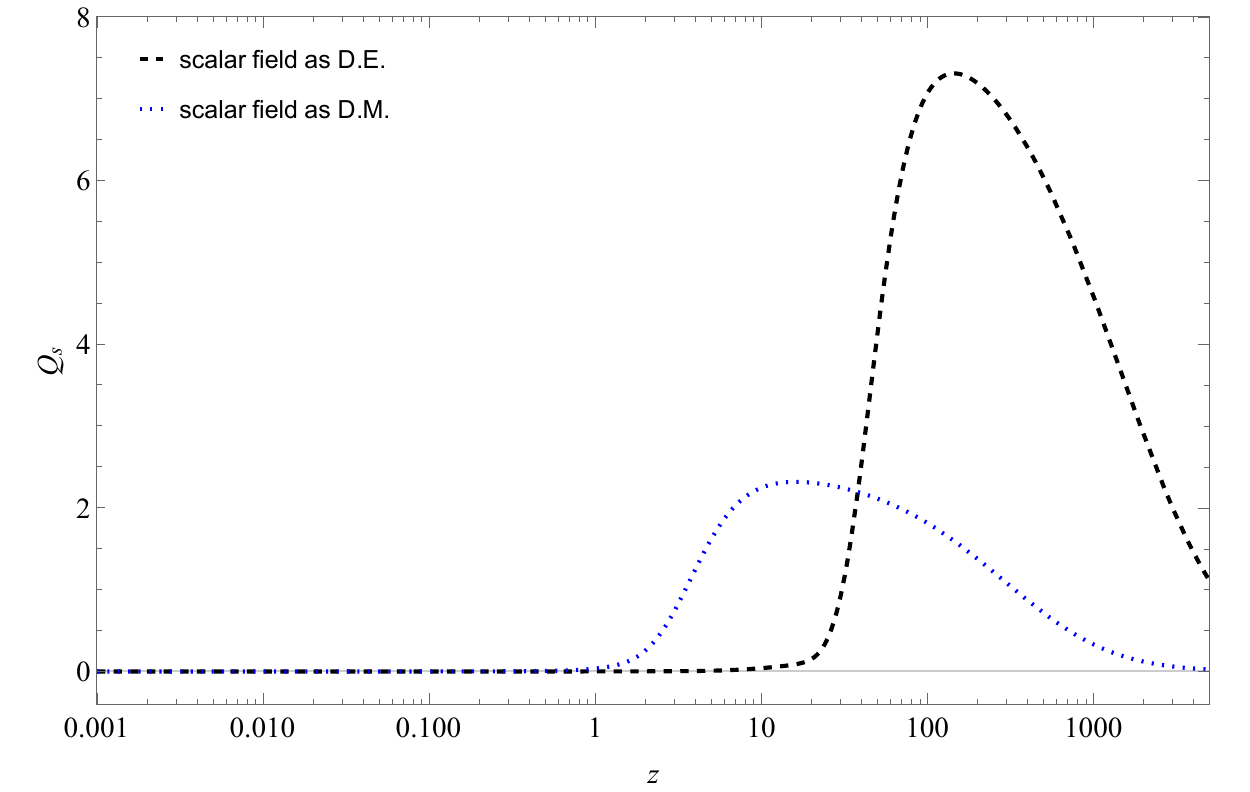}\label{grafQS}}
  \caption{\footnotesize{In Fig. \ref{grafcS}, we show the evolution of the squared propagation speed of scalar perturbations, $c_S^2$, as a function of redshift $z$. In Fig. \ref{grafQS}, we present the evolution of the kinetic energy parameter associated with scalar perturbations, $Q_S$, also as a function of redshift $z$. In both cases we have the contribution of the scalar field as dark matter (dotted blue lines) and the scalar field as dark energy (dashed black lines)}}
 \label{PLOcSQS}
\end{figure}

Next we write the equations of the parameters of scalar and tensor perturbations expressed in \eqref{eqcs2}, \eqref{eqcT} and \eqref{eqqs} in terms of the new $w_i$ of the model. 
In doing this, we examine the stability of the solutions obtained for the model, investigating the square of the velocities of the scalar and tensor perturbations, in addition to the kinetic energy associated with the scalar $Q_S$ and tensor $Q_T$ perturbations. Thus, for consistent dynamics, free of Laplacian and ghost instabilities for the scalar and tensor modes, the following conditions must be satisfied,
\begin{equation} \label{eqcond}
    c_S^2\geq0, \qquad Q_S>0, \qquad  c_T^2 \geq 0\qquad \text{and} \qquad Q_T>0.
\end{equation}
In this way, we analyze the evolutions of \eqref{eqcs2}, \eqref{eqqs} and \eqref{eqcT} as a function of redshift for the background solution given by the model of the equations \eqref{eqmodel}. 
In Fig. \ref{PLOcSQS}, we evolve the propagation velocity (left panel) and kinetic energy (right panel) associated with the scalar perturbations. In the figures in this section, we use the best fit results for the combination CC$+$BAO$+$SH$0$ES data.
For $Q_{s}$, the stability conditions are satisfied throughout the evaluated interval, showing that the solutions obtained for the model are free of ghost instabilities. When we investigate the behavior of $c_{S}$, the behavior of the curves admits Laplacian stability within the analyzed redshift range $0<z<5000$ (This upper bound redshift is far beyond typical astronomical observations). As we can see in the highlighted boxes in Fig. \ref{grafcS}, The curves remain above zero for both case: the scalar field playing the role of dark energy (dashed black line) and dark matter (dotted blue line). This is precisely the regime where the Horndeski gravity is completely safe, as shown in the extension of the theory presented in Ref.~\cite{Gleyzes:2014dya}.
In particular, in both cases we observe that at a certain late time epoch the sound-speed squared approaches zero, reaching $c_{s}^{2} \approx 7.8\times 10^{-3}$ at $z \approx 2.4$ for the scalar field as dark matter and $c_{s}^{2} = 1.9 \times 10^{-4}$ at $z \approx 30.5$ for dark energy. The regime of reduced speed of sound is confined to a short finite redshift interval for both cases.

\begin{figure}[htbp]
 \centering
\includegraphics[width=.5\textwidth]{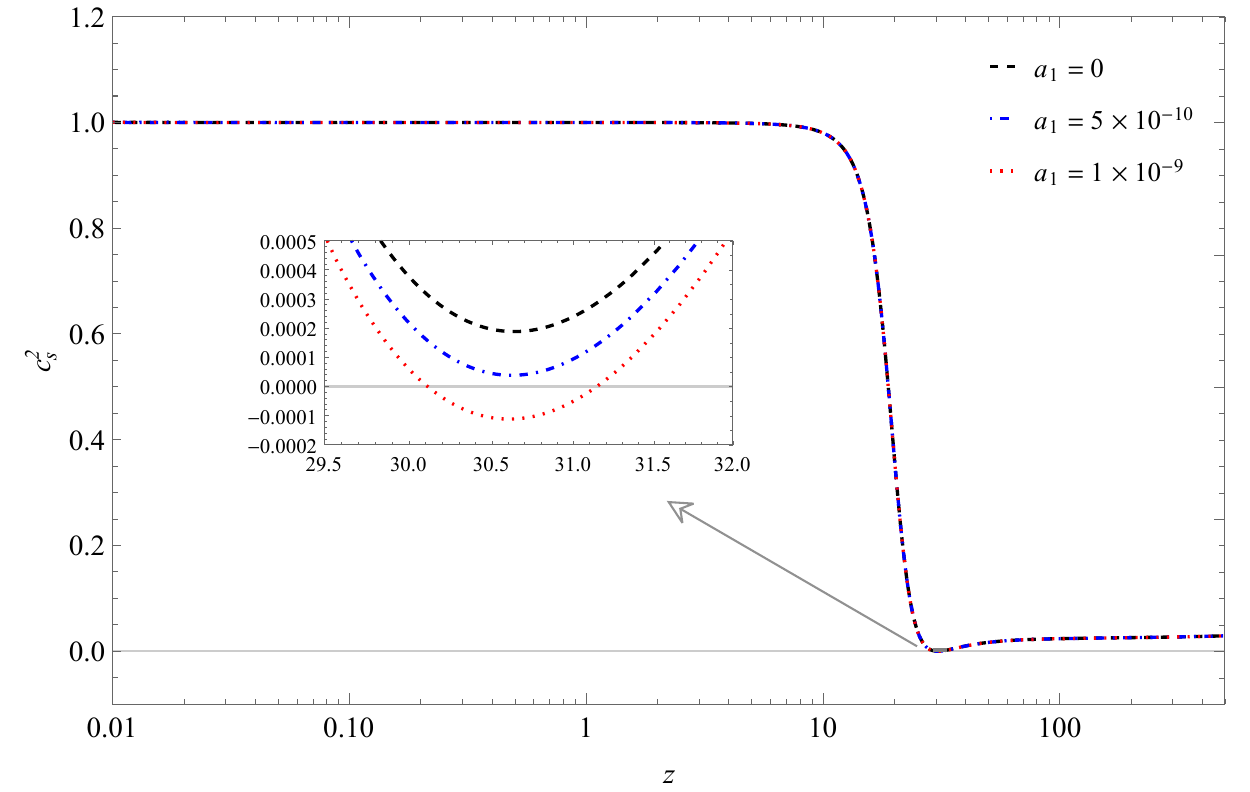}
\caption{\footnotesize{The figure shows the evolution of the speed of sound squared, $c_{s}^{2}$, of scalar perturbations for the scalar field model as dark energy, including the contribution of the Horndeski cubic $G_{3}(X,\phi)$. The inset highlights how different values of the parameter $a_{1}$ (associated with $G_{3}(X,\phi)$) affect $c_{s}^{2}$.}}
 \label{grafcSG3}
\end{figure}

It is important to highlight that, in Horndeski theories, cubic-order interaction terms in scalar perturbations contain inverse powers of the speed of sound $c_{s}$. Then, when $c_{s}<<1$, these terms can become dominant, calling into question the validity of perturbative analysis based solely on the quadratic Lagrangian. Thus, even if the speed of sound is non-zero, a very small $c_{s}$ can significantly reduce the effective field theory (EFT) limit and may lead to the breaking of perturbative unitarity for the EFT of cosmological perturbations~\cite{de2011generalized, pirtskhalava2015weakly, santoni2018behind}.
To explore this issue, we can perform a preliminary analysis to verify the effects of a specific cubic term $G_{3}(X,\phi) = a_{1}X e^{\lambda\phi} + a_{2}X^{2} e^{2\lambda\phi}$~\cite{PhysRevD.98.064038} on equations \eqref{eqJ} and \eqref{eqPf}, in the intermediate energy regime,
\begin{eqnarray}
  J &=& \left(\alpha+3\eta H(t)^2\right)\dot{\phi}(t) + 3 a_{1}H(t)\dot{\phi}(t)^2,\\
  P_{\phi} &=&-\lambda e^{\lambda\phi(t)}\left(V_{0}+a_{1}\dot{\phi}(t)^{2}\ddot{\phi}(t)\right),
\end{eqnarray}
where in this approximation we kept up to quadratic terms in $\dot{\phi}(t)$.
For this scenario, using the same parameter values as in the dark energy case Fig. \ref{PLOcSQS}, we compute the resulting squared speed of sound $c_s^{2}$. The results with the contribution of $G_{3}(X,\phi)$ are shown in Fig. \ref{grafcSG3}. 
We find that for sufficiently small value of the parameter $a_{1}$ derived from $G_{3}(X,\phi)$, the square of the speed of sound remains positive, as can be seen in the inset of the figure.
This is an approximate view of the issue aforementioned. 
We believe that further considerations should be addressed everywhere.

One of the recurring discussions within the framework of Horndeski theories concerns the possibility of a time variation in the speed of gravitational waves in models that involve the terms $G_4(\phi, X)$ and $G_5(\phi, X)$. 
In this context, the emergence of superluminality does not necessarily violate causality, as discussed by several authors~\cite{babichev2008k, ellis2007causality, eperon2019predictability}. Fundamental causality is defined by light cone of the physical spacetime metric. However, for theories such as Horndeski, this concept does not apply directly, since the propagation of perturbations is governed by an ``effective light cone'', which emerges from the background field dynamics.
The observation of gravitational waves GW170817 highlighted the relevance of these effective metrics. As shown by Ezquiaga and Zumalacárregui~\cite{ezquiaga2017dark}, this event restricted the speed of gravitational waves to values very close to the speed of light, $c_{T} \approx 1$, ruling out a wide class of Horndeski models with a non-trivial evolution of the $G_5(\phi, X)$ and $G_4(\phi, X)$ functions. For the scalar field as dark energy, our model escapes this strict restriction, since $G_5(\phi, X)$ is approximately constant. A constant $G_5(\phi, X)$ function does not contribute to the deviation of $c_{T}$ from unity, since the relevant terms in the perturbation analysis involve its derivatives, allowing the scalar field to act as dark energy without contradicting the gravitational wave observations.
As discussed by Casalino et al~\cite{CASALINO2019100243}, to obtain a dark matter sector from the scalar field requires a non-minimal coupling between the scalar field and the curvature. This coupling generates the derivation $c_{T} \neq 1$. It is crucial to note, however, that the gravitational waves propagate in an effective medium constituted by the dark matter field itself, and not in a vacuum. 

\begin{figure}[htbp]
 \centering
\subfigure[]{\includegraphics[width=.45\textwidth]{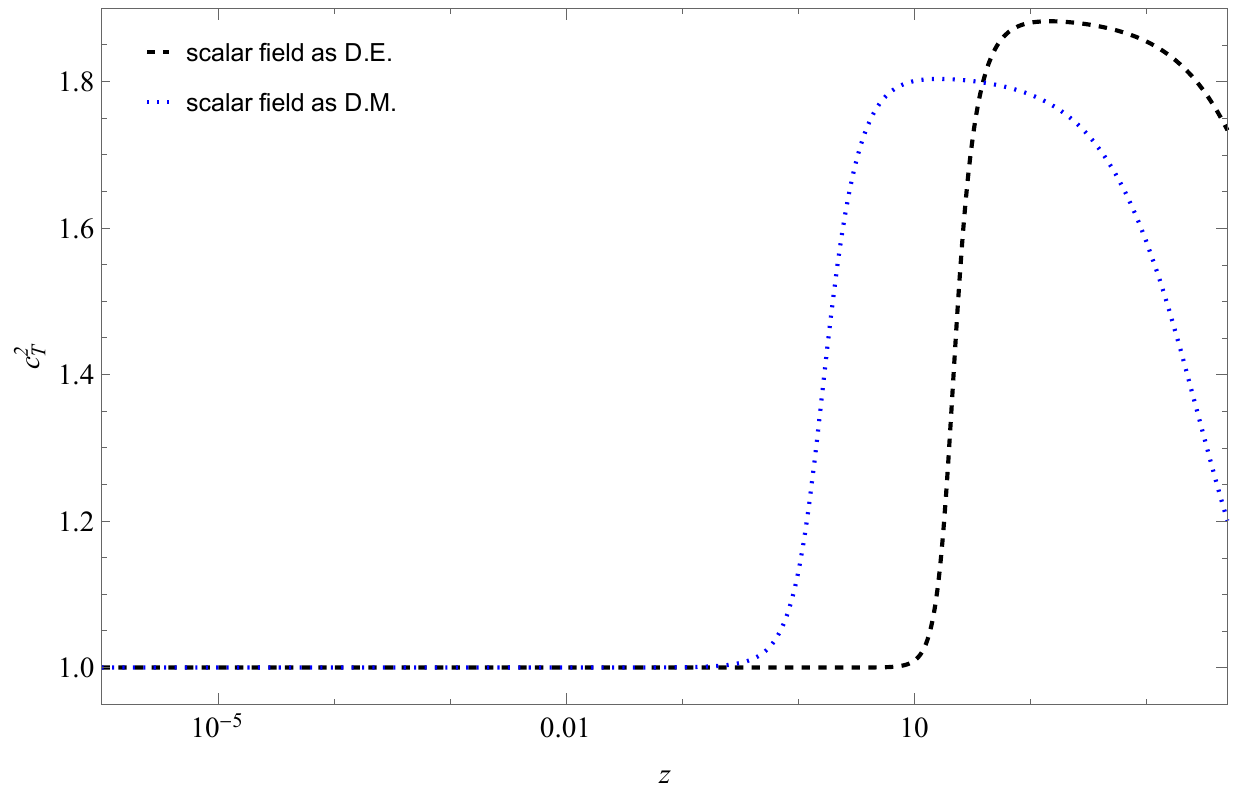}\label{grafcT}}
 \quad
 \subfigure[]{\includegraphics[width=.45\textwidth]{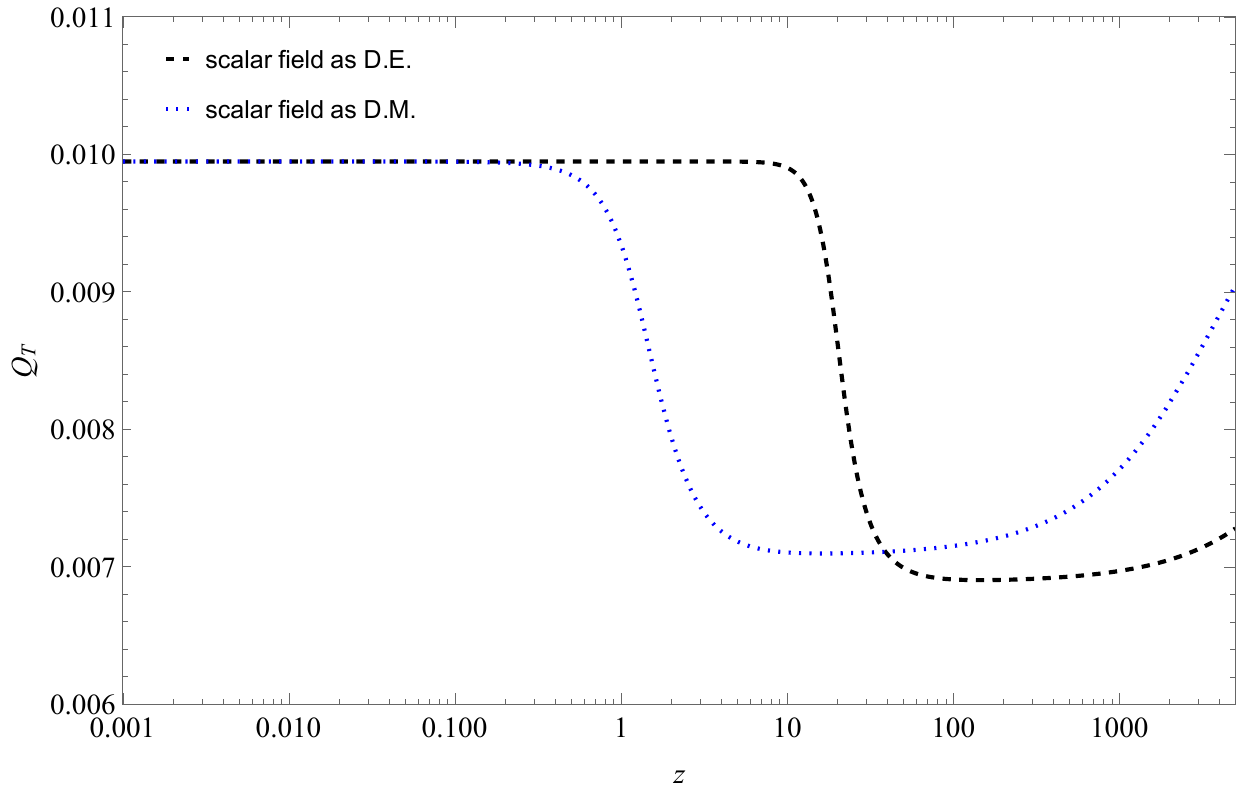}\label{grafQT}}
 \caption{\footnotesize{In Fig. \ref{grafcT}, we show the evolution of the squared propagation speed of tensor perturbations, $c_T^2$, as a function of redshift $z$. At $z=0$ we obtain:$|c^{2}_{T}-1| \approx 0.4\times 10^{-16}$ for D.E. and $|c^{2}_{T}-1| \approx 1.1\times 10^{-15}$ for D.M.. Meanwhile, in Fig. \ref{grafQT}, we present the evolution of the kinetic energy parameter associated with tensor perturbations, $Q_T$, as a function of redshift $z$.}}
 \label{PLOcTQT}
\end{figure}

In our study, we consider $G_5(\phi, X) = -\eta \phi/2$, which, in principle, does not guarantee that the speed of gravitational waves is exactly luminal ($c_T^2 = 1$). For this reason, contributions from $G_5(\phi, X)$ are commonly neglected in many approaches based on this class of theories, as previously discussed. However, in the model that we are analyzing, by properly adjusting the free parameters and making appropriate initial conditions choices, it is possible to obtain a velocity $c_T^2 \approx 1$ at $z \approx 0$, while remaining positive throughout the entire evolution interval considered. This behavior can be verified in the graph on the left of Fig. \ref{PLOcTQT}. Furthermore, the constraint \eqref{ct_cond} can be verified using Eq. \eqref{eqcT} in the $z=0$ regime. At this redshift, the propagation speed can be written in the form:
\begin{equation}
    c_{T}^{2} = \dfrac{4\kappa + \eta H_{0}^{2}\phi'(z=0)^{2}}{4\kappa - \eta H_{0}^{2}\phi'(z=0)^{2}}.
    \label{ctz0}
\end{equation}
Imposing the constraint \eqref{ct_cond} on Eq. \eqref{ctz0} with the initial value of $\phi'(z=0) = 5 \times 10^{-7} M_{\mathrm{Pl}}$, we get the constraint on the constant $\eta$
\begin{equation}
    \eta H_{0}^{2} \le 1.59\times10^{-4}.
\end{equation}
Consequently, the best fit results from the combined CC$+$BAO$+$SH$0$ES data (see Tables \ref{tabss} and \ref{tabbf2}) yield the following values: $\eta H_{0}^{2} = \left(0.53\pm0.08\right)\times10^{-4}$ for the scalar field as dark energy and $\eta H_{0}^{2} = \left(1.63\pm 0.04\right)\times10^{-4}$ for the dark matter. For the scalar field as dark energy, the result is within the limit, satisfying the constraint. However, for the dark matter scenario, the best fit value shows a slight discrepancy from our theoretical limit, satisfying only the lower part of the uncertainty. Using the CC data from Table \ref{tabss} we obtain $\eta H_{0}^{2} = \left(1.42\pm 0.05\right)\times10^{-4}$ for dark matter; this result satisfies the constraint. 

\begin{figure}[htbp]
 \centering
\subfigure[]{\includegraphics[width=.45\textwidth]{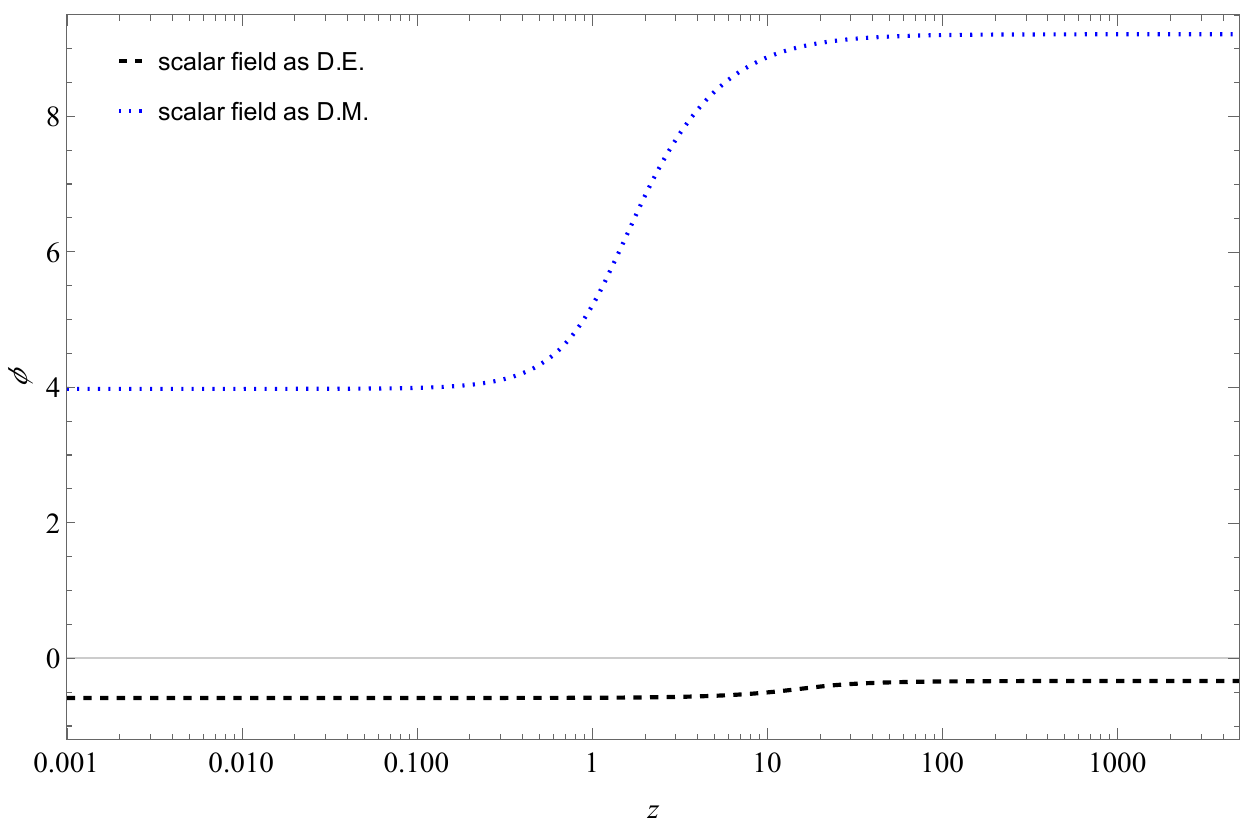}\label{grafphi}}
 \quad
 \subfigure[]{\includegraphics[width=.45\textwidth]{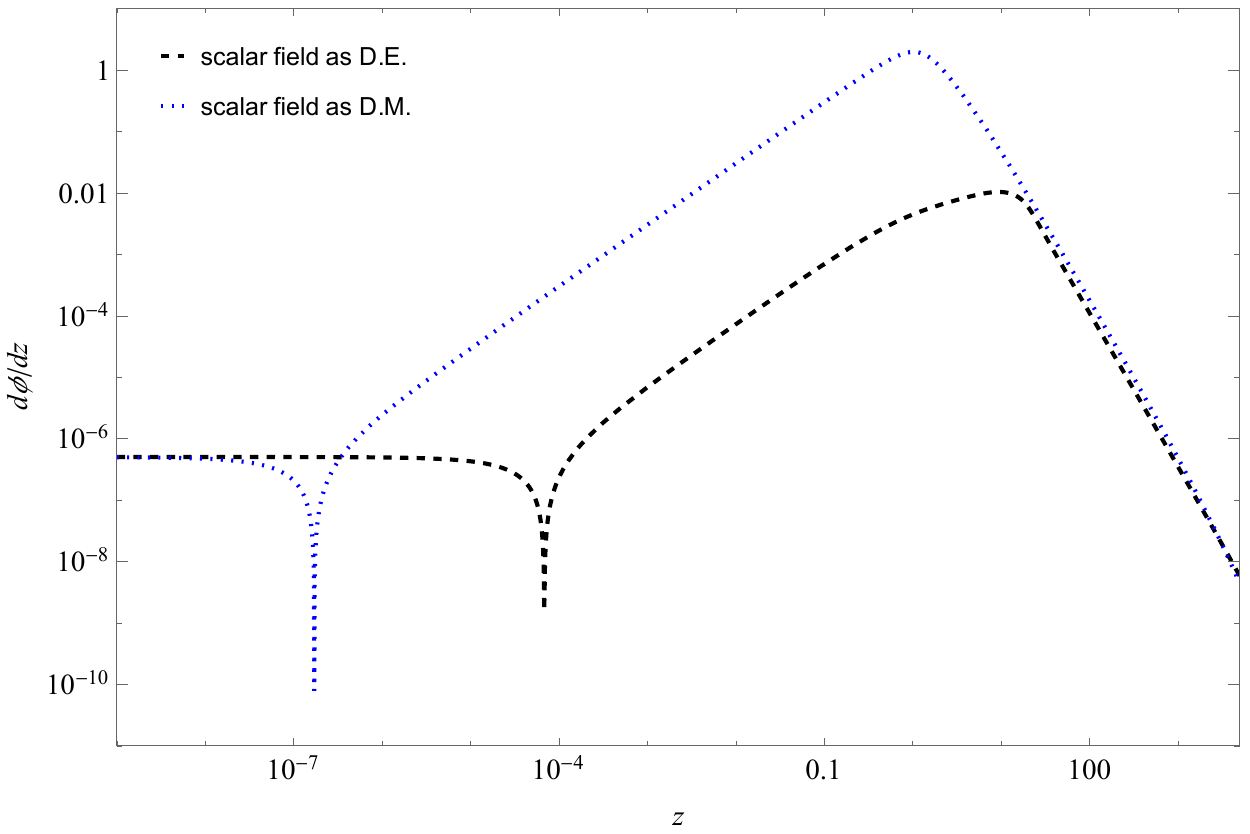}\label{grafdphi}}
 \caption{\footnotesize{Evolution of the scalar field $\phi$ and its derivative with respect to redshift, using the best-fit results for the CC$+$BAO$+$SH$0$ES data.The right panel shows the behavior of the derivative of $\phi(z)$ in a Log-Log plot.}}
 \label{Plotphi}
\end{figure}

Regarding the kinetic energy quantity $Q_T$, shown on the right of Fig. \ref{PLOcTQT}, it also remains positive throughout the entire evolution. These results ensure that the model satisfies the stability conditions and is, therefore, free from ghost and Laplacian instabilities associated with tensor modes. In Fig.\ref{Plotphi} we have the behavior of the scalar field $\phi(z)$ and its derivative $\phi'(z)$ for both models. We can see that for the scalar field acting as dark energy, $\phi$ exhibits an almost constant behavior. This behavior suggests that for dark energy, $G_{5}(\phi)$ is approximately constant. We can also study an important cosmological quantity, the parameter associated with the {\it effective equation of state} for the scalar field given by $\omega_{\phi} = p_{\phi}/\rho_{\phi}$, which can be obtained by using the equations \eqref{eqdensidphi} and \eqref{eqpphi}. 

 \begin{figure}[htbp]
 \centering
 \includegraphics[width=.5\textwidth]{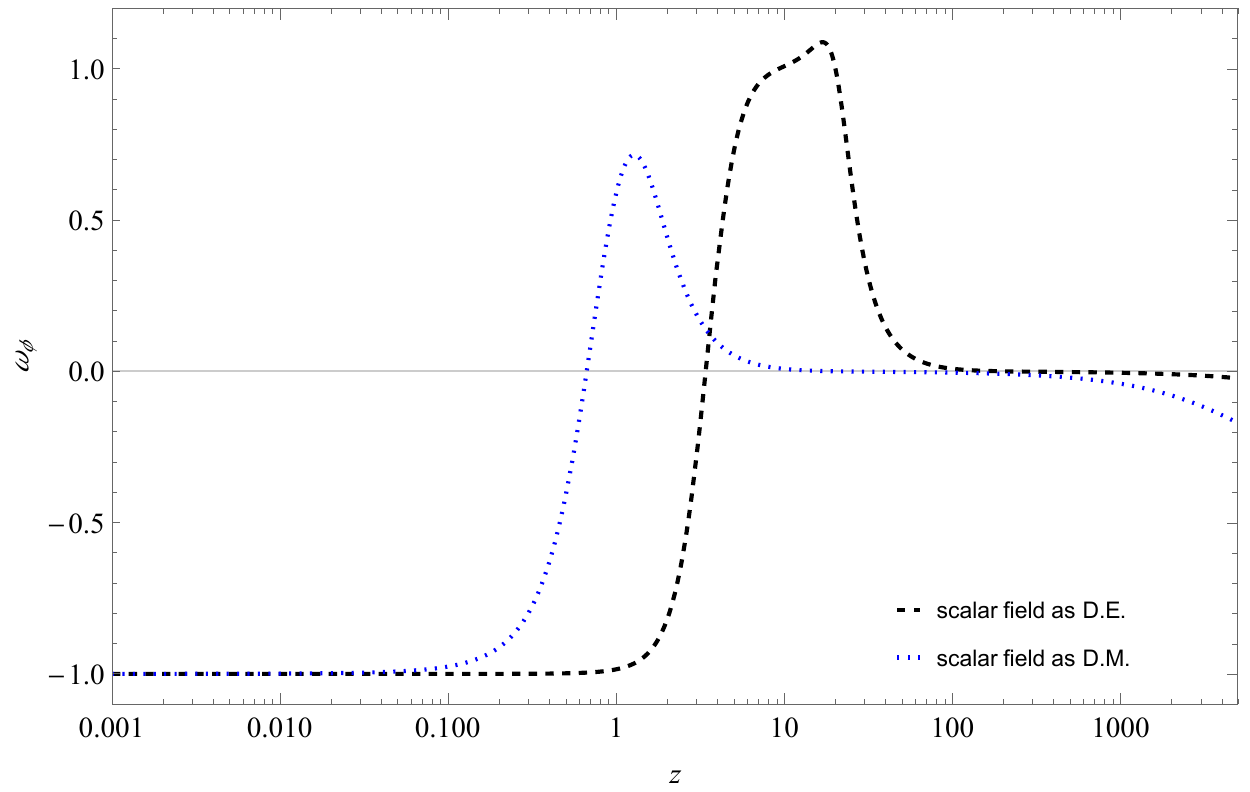}
  \caption{\footnotesize{Evolution of the effective equation of state associated with the scalar field $\omega_{\phi}$ as a function of redshift $z$ for the scalar field as dark energy (dashed black line) and the scalar as dark matter (dotted blue line).}}
 \label{figEqState}
\end{figure}
We can see in Fig.~\ref{figEqState} the behavior of the equation of state for both cases using the best-fit results for the CC$+$BAO$+$SH$0$ES data. For the scalar field as dark energy (dashed black line), we have a slight peak where $\omega_{\phi}>1$ (`super stiff matter' regime), reflecting the fact that $\omega_\phi$ is indeed an effective equation of state.

\section{Conclusions}
\label{sec6}

In this work, we explore the effects of Horndeski gravity integrated with the theoretical and observational foundations of the standard cosmological model.
We analyze the statistics, using the observational data of $H(z)$ from the SH$0$ES, BAO, and CC datasets, applied to a specific model featuring a non-minimal derivative coupling between the scalar field and Einstein tensor. Two scenarios were investigated: in the first (Case I), the scalar field acts as dark matter; in the second (Case II), the scalar field is responsible for the dark energy component of the Universe, replacing the cosmological constant in driving cosmic expansion.

This approach allowed us to obtain estimates for the values of the main cosmological parameter, as can be seen in the posterior distributions and in the tables. 
In this way, limits were set for the coupling parameters $\alpha$ and $\eta$ of the model, yielding values that fit well within the current cosmological framework. Particular attention was paid to the $H_0$ values obtained in both cases. We found that using different $H(z)$ datasets, as well as their combinations, resulted in Hubble constant values within the ranges reported by P20 and R22. This result is particularly relevant, as it contributes to alleviating the tension between local and global determinations of $H_0$. Additionally, we observed that the behavior of $H(z)$ remains within the expected range when evolved over $0 \leq z \leq 2.5$.

In Case I the best mitigation of the tension was achieved using combined CC$+$BAO$+$SH$0$ES data, reducing the tension with R22 to only $0.38\sigma$, while in Case II, the reduction reached $1.7\sigma$ using the combined data, which provides relief of the tension with R22. Overall, the model studied, which has been extensively investigated in recent literature in various contexts of gravity and cosmology proved to be both promising and consistent with the results of our analysis.

 Finally, concerning the effective equation of state, in neither the aforementioned case, was the scalar field able to cross the phantom divide at large redshifts as recently pointed and explored by the DESI collaboration \cite{DESI:2025zgx,DESI:2025fii}. This is a point to be explored in our setup in the realm of the DESI dataset in upcoming investigations, along the lines of \cite{daCosta:2024grm,Akrami:2025zlb,Ye:2025ulq} in the context of ``beyond Horndeski'' physics.

{\acknowledgments We thank CNPq and CAPES
for partial financial support. FAB acknowledges
support from CNPq (Grant No. $309092/2022-1$). JAVC thanks the Paraíba State Research Support Foundation (FAPESQ) (Grant No. $22/2025$) for financial support. The authors thank Amilcar R. Queiroz for his helpful comments.
\\
\\
\centerline{{\it In memory} of our late collaborator and friend Prof. Raimundo Silva.} 
}

\end{document}